\documentclass[12pt]{article}
\usepackage{amssymb,amsmath,bm,bbm}
\usepackage{epsf}
\usepackage{epsfig}
\usepackage{afterpage}
\usepackage{longtable}
\usepackage{cite}
\usepackage{latexsym, mathrsfs}
\usepackage{axodraw}
\usepackage{graphics}
\usepackage{color}

\def\slash#1{\setbox0=\hbox{$#1$}\dimen0=\wd0
      \setbox1=\hbox{/} \dimen1=\wd1 \ifdim\dimen0>\dimen1
      \rlap{\hbox to \dimen0{\hfil/\hfil}} #1                        \else
      \rlap{\hbox to \dimen1{\hfil$#1$\hfil}}
      /   \fi}

\newcommand{\lsim}{
\mathrel{\hbox{\rlap{\hbox{\lower4pt\hbox{$\sim$}}}\hbox{$<$}}}}

\newcommand{\gsim}{
\mathrel{\hbox{\rlap{\hbox{\lower4pt\hbox{$\sim$}}}\hbox{$>$}}}}

\newcommand{\vud}{V_{ud}}
\newcommand{\vcd}{V_{cd}}
\newcommand{\vtb}{V_{tb}}
\newcommand{\vus}{V_{us}}
\newcommand{\vcb}{V_{cb}}
\newcommand{\vtd}{V_{td}}
\newcommand{\vub}{V_{ub}}
\newcommand{\vts}{V_{ts}}
\newcommand{\vcs}{V_{cs}}

\newcommand{\gev}{\, {\rm GeV}}

\def\beq{\begin{equation}}
\def\eeq{\end{equation}}

\newcommand{\be}{\begin{equation}}
\newcommand{\ee}{\end{equation}}
\newcommand{\bea}{\begin{eqnarray}}
\newcommand{\eea}{\end{eqnarray}}

\newcommand{\bi}{\begin{itemize}}
\newcommand{\ei}{\end{itemize}}

\newcommand{\newsection}[1]{\section{#1}\setcounter{equation}{0}}

\definecolor{gray}{rgb}{.38,.38,.38}

\begin{document}
\begin{titlepage}
\vspace*{-0.5truecm}

\begin{flushright}
{TUM-HEP-715/09}\\
{MPP-2009-46}
\end{flushright}

\vfill

\begin{center}
\boldmath

{\Large\textbf{The Impact of Kaluza-Klein Fermions on\\ Standard Model Fermion Couplings in a RS Model with Custodial Protection}}

\unboldmath
\end{center}

\vspace{0.3truecm}

\begin{center}
{\bf Andrzej J.~Buras$^{a,b}$, Bjoern Duling$^a$ and Stefania Gori$^{a,c}$}
\vspace{0.4truecm}

{\footnotesize
 $^a${\sl Physik Department, Technische Universit\"at M\"unchen,
D-85748 Garching, Germany}\vspace{0.2cm}

 $^b${\sl TUM Institute for Advanced Study, Technische Universit\"at M\"unchen,  \\ D-80333 M\"unchen, Germany}\vspace{0.2cm}

 $^c${\sl Max-Planck-Institut f{\"u}r Physik (Werner-Heisenberg-Institut), \\
D-80805 M{\"u}nchen, Germany}\vspace{0.2cm}
}

\end{center}

\begin{abstract}

\noindent
We reconsider the impact of heavy vector-like fermions on the couplings of
standard model (SM) quarks to the SM gauge bosons $W^\pm$ and $Z$ and to the
SM Higgs boson $H$. Integrating out these fermions at tree level we derive general
formulae that can be used in any model containing such particles. We apply
these formulae to the case of the lightest Kaluza-Klein (KK) fermions in a Randall-Sundrum
(RS) model with a custodial protection of flavour conserving $Zd_L^i\bar d_L^i$
and flavour violating $Zd_L^i\bar d_L^j$ couplings. We point out that in this
model also the couplings of $Zu_R^i\bar u_R^i$ and $Zu_R^i\bar u_R^j$ are
protected. In particular we demonstrate explicitly that this protection is not
spoiled by the mixing of the SM quarks with the KK fermions, which is
guaranteed by the underlying $P_{LR}$ symmetry. We find that the impact of KK
fermions on the $Z$-couplings, while not negligible, is significantly smaller than the one coming from the mixing of $Z$ with
the heavy KK gauge bosons $Z_H$ and $Z'$. {The situation is similar for the $W^\pm$ couplings with the only exception of the $tb$ coupling where the impact of KK fermions can in principle be larger than the effects that are induced by gauge boson mixing.} We also show explicitly that at $\mathcal{O}(v^2/M_\text{KK}^2)$ the
fermion--Higgs couplings of a Higgs placed on the infrared (IR) brane is not
affected by the KK contributions up to strongly chirally suppressed
contributions.
 The corrections to the CKM matrix, in particular the breakdown of its unitarity,
 are found to be small. We also investigate the
 right-handed couplings of $W^\pm$ that are generated through the mixing with
 KK-fermions and introduce a $3\times 3$ matrix that describes the pattern of flavour
 violation in these new interactions of the standard $W^\pm$.
\end{abstract}

\end{titlepage}

\setcounter{page}{1}
\pagenumbering{arabic}

\newsection{Introduction}
There are many extensions of the SM in which new fermions with non-standard $SU(2)_L\times U(1)_Y$ quantum numbers are present. In this context a prominent role is played by vector-like fermions, that is fermions  whose left-handed and right-handed components transform in the same way under $SU(3)_c\times SU(2)_L\times U(1)_Y$. As such, their masses are not protected by the SM gauge symmetry and can be much larger than the masses of the ordinary quarks. Consequently, the effects of such fermions on electroweak observables can be put easier under control than those coming e.g.~from the fourth generation of ordinary chiral fermions.

On the other hand the mixing of the vector-like fermions with the ordinary quarks, which is caused by Yukawa couplings, implies necessarily the breakdown of the GIM mechanism and the appearance of flavour changing neutral current (FCNC) transitions already at the tree level. As these transitions are very strongly suppressed in nature, the presence of FCNC processes in a given model at tree level could be problematic unless the new physics (NP) scale is much larger than 1~TeV or there is a protection mechanism for the FCNC couplings.
This is in particular the case for processes involving the external down-quarks, like $K^0-\bar K^0$, $B_d^0-\bar B_d^0$, $B_s^0-\bar B_s^0$ mixings and rare $K$ and $B$ decays for which a large amount of data already exists. On the other hand much less is known about the FCNC processes involving up-quarks, but this situation will clear up in the LHC era.

Models with vector-like quarks have been considered extensively in the
literature. An excellent presentation of these models and of the machinery
involved can be found in \cite{Branco:1999fs}. Moreover, del Aguila and
collaborators analysed in a series of papers 
\cite{delAguila:2000kb,delAguila:2000aa,delAguila:2000rc} the mixing of vector-like quarks with ordinary quarks in the effective Lagrangian approach as was performed by Buchm\"uller and Wyler in a more general analysis~\cite{Buchmuller:1985jz}. Integrating out these new heavy fermions by means of equations of motion (EOM) at the tree level, {del Aguila et al.~}derived general expressions for the corrections to the $Vq\bar q^\prime$ ($V=Z,W^\pm$) and $Hq\bar q^\prime$ couplings that come from the mixing of heavy and ordinary quarks. In particular in \cite{delAguila:2000kb} del Aguila and Santiago analysed the impact of KK fermion excitations on $Vq\bar q^\prime$ couplings in a RS model.

Now in RS models there is another mechanism for generating tree level FCNC
transitions which is connected with the gauge sector. Indeed in these models
the interactions of ordinary quarks with the KK gauge bosons are non-universal
in flavour and this non-universality implies tree level FCNC processes
mediated by these heavy gauge bosons. Moreover, due to the mixing of these
heavy gauge bosons with the SM $Z$ boson in the process of electroweak
symmetry breaking (EWSB), also the $Z$ couplings to quarks become flavour
non-universal implying in turn tree level FCNC processes mediated by the $Z$ boson. 
{Also corrections to the CKM matrix 
including the violation of its unitarity come both from the mixing in the
gauge sector in question and from the mixing with KK fermions.}

In two recent papers \cite{Blanke:2008zb,Blanke:2008yr} we have analysed particle-antiparticle mixing and rare decays of $K$ and $B$ mesons in a particular RS model with an extended gauge group
\begin{equation}\label{eq:gaugegroup}
SU(3)_c\times SU(2)_L\times SU(2)_R\times U(1)_X\times P_{LR}\,.
\end{equation}
Thanks to the symmetries $SU(2)_R$ and $P_{LR}$ in this model the $T$
parameter \cite{Agashe:2003zs,Csaki:2003zu} and the coupling $Zb_L\bar b_L$ \cite{Agashe:2006at} are protected from new tree level contributions up to the small breaking of the $P_{LR}$ symmetry on the UV brane. This allows to satisfy the very stringent electroweak constraints with KK scales of order $(2-3)$TeV which are in the reach of the LHC. In \cite{Blanke:2008zb} we have pointed out that the custodial symmetry $P_{LR}$ {together with appropriate fermion representations} automatically implies the protection of flavour violating $Zd_L^i\bar d_L^j$ couplings so that tree level $Z$ contributions to all processes in which the flavour changes appear in the down quark sector are dominantly represented by $Zd_R^i\bar d_R^j$ couplings. Moreover, we found in \cite{Blanke:2008yr} that, for $\Delta F=1$ observables, these exchanges dominate over the exchanges of heavy $Z_H$ and $Z'$ present in this model. This has profound implications for the pattern of flavour violation in rare $K$ and $B$ decays as discussed in detail in \cite{Blanke:2008yr}.\\

Our analysis in \cite{Blanke:2008zb,Blanke:2008yr} was performed in the
approximation of neglecting the mixing of ordinary quarks with the KK
fermions, which, as discussed above, is another source of tree level FCNC
transitions in addition to the one from the gauge sector. 

{Our detailed analysis in the context of the RS model with custodial protection in question shows that
the impact of the mixing in the gauge sector on the SM gauge couplings to quarks is definitely larger than the one coming from KK mixing. Only in the case of charged left-handed currents involving the top quark, KK mixing can compete 
with the mixing in the gauge sector and it is the sole effect in this 
model that is responsible for the generation of the right-handed couplings of $W^\pm$.
In any case, in processes involving external down quarks, the effects of mixing between ordinary quarks and KK fermions are subleading and the approximation of neglecting these contributions made in \cite{Blanke:2008zb,Blanke:2008yr} seems to be justified.}

However, as we have only considered the effects of the lightest KK fermions it would be of interest to see whether the summation of all KK fermion contribution would significantly modify our findings\footnote{The summation over all KK fermion contributions was found to have a significant effect in the case of flavour changing couplings of the Higgs boson\cite{Azatov:2009na}}. We postpone this to a future analysis.\\

In our presentation it turned out to be useful not to use the general
expressions for the modified SM couplings in \cite{delAguila:2000kb}, but
rather derive the corrections from mixing with KK fermions directly. This in
turn allowed us to discuss the impact of custodial protection on this mixing
more transparently than in our view could be done with the general formulae in
\cite{delAguila:2000kb}.\footnote{We have been informed by the authors of \cite{delAguila:2000kb,delAguila:2000aa,delAguila:2000rc} that they confirmed
our results
for the effective quark couplings of the $Z$ boson
using their approach.}

 In their original paper, del Aguila et al.~found the effects from KK fermion mixing to be potentially important, in particular in the so-called ``conformal window'', while a comparison with gauge boson mixing effects was not performed in their paper. Our present work goes beyond their analysis in three ways. First, we perform our analysis in the phenomenologically favoured RS model with custodial protection, being able to show analytically how the cancellations implied by the custodial symmetry occur. Second, having at hand the results of \cite{Blanke:2008zb,Blanke:2008yr} we perform a detailed comparison of the effects attributed to KK fermion mixing and gauge boson mixing. Finally, again owing to our previous analyses we are in the position to perform this comparison for a large set of actual parameter points that are chosen to reproduce the SM masses and mixings and beyond that are consistent with the available experimental constraints on $\Delta F=2$ and $\Delta F=1$ observables.\\

Our paper is organised as follows. In Section~2 we present the fundamental
Lagrangian involving new vector-like quarks in addition to the SM quarks in a
form suitable for the main goals of our paper. In Section~3 we integrate out
the heavy fermions by means of EOM. The basic result of this
section are the master formulae that express the heavy quark fields in terms
of the ordinary quark fields and the formulae summarising the shift in the
light quark fields necessary to bring their kinetic terms into canonical
form. However, the main formulae of our paper are collected in Section~4,
where we give general expressions for the corrections to $Z$, $W^\pm$ and $H$
couplings resulting from the mixing with heavy fermions that have been
integrated out. We would like to emphasise that these formulae are general and
can be applied to any extension of the SM that contains vector-like
fermions. In Section~5 we review the custodial protection of neutral $Z$ and $Z^\prime$ couplings in the context of the RS model analysed in detail in \cite{Blanke:2008zb,Blanke:2008yr,Albrecht:2009xr}. We generalise the discussion of these papers to SM up-quarks couplings and note the protection of certain couplings involving KK-fermions.

In Section 6 we apply the formalism of Sections 2-4 to analyse the impact of
KK-fermions on the SM neutral couplings, that is of $Z$ and $H$, in the
context of the RS model in question. Here we demonstrate explicitly that the
custodial protection of Section~5 remains intact in the presence of
KK-fermions as expected on the basis of symmetry arguments. In Section~7 we
analyse the impact of mixing in the gauge boson sector and of the KK-fermions
on the SM couplings of $W^\pm$. In particular we discuss the violation of unitarity of the CKM matrix and the corrections to its elements. We also
investigate the $W^\pm$ couplings to right-handed quarks that originate from
the mixing of the SM quarks with vector-like KK-fermions. 

In Section~8 the numerical analysis of all topics discussed in Sections~5-7 is presented. Moreover, we analyse the accuracy of our formulae of Section~4 that include only $\mathcal{O}(v^2/M_\text{KK}^2$) corrections numerically. Here we also confront the size of the corrections to the SM couplings from gauge boson mixing with the corrections from KK-mixing finding the former dominant. We summarise the main results of our paper in Section~9. In the Appendix we collect the couplings and charge factors that we use in Sections~5-7.\footnote{It would be of interest to see how our results would be changed in a RS model without custodial protection as studied by the authors of~\cite{Casagrande:2008hr,Bauer:2008xb}} 

\newsection{Fundamental Lagrangian}\label{sec:fundamental-lagrangian}
\subsection{Preliminaries}
We consider a theory which in addition to SM quarks contains $N$ charge $+2/3$ heavy vectorial fermions and $M$ charge $-1/3$ heavy vectorial fermions\footnote{{The following discussion is not affected by} the presence of additional fermions with a different charge than those of the SM {as is the case for example in the RS model considered in~\cite{Blanke:2008zb,Blanke:2008yr,Albrecht:2009xr}}.}.
Similarly to quarks these fermions are put in triplet representations under $SU(3)_c$ and as quarks each of them appears in three flavours. Suppressing the colour and flavour indices for the moment we introduce the vectors
\begin{eqnarray}
\Psi_L^T(2/3)&=&\left(u_L,\,U_L^1,\,U_L^2,\dots ,\,U_L^N\right)\label{eq:2.1}\,,\\
\Psi_R^T(2/3)&=&\left(u_R,\,U_R^1,\,U_R^2,\dots ,\,U_R^N\right)\label{eq:2.2}\,,\\
\Psi_L^T(-1/3)&=&\left(d_L,\,D_L^1,\,D_L^2,\dots ,\,D_L^M\right)\label{eq:2.3}\,,\\
\Psi_R^T(-1/3)&=&\left(d_R,\,D_R^1,\,D_R^2,\dots ,\,D_R^M\right)\label{eq:2.4}\,,
\end{eqnarray}
with $u_{L,R}$ and $d_{L,R}$ denoting SM quarks and $U_{L,R}^i$ and
$D_{L,R}^i$ heavy quarks. All the entries in (\ref{eq:2.1})-(\ref{eq:2.4}) are three dimensional row vectors in the flavour space so that $\Psi_{L,R}(2/3)$ have $3(N+1)$ components while $\Psi_{L,R}(-1/3)$ have $3(M+1)$ components. Colour indices will be always suppressed, while flavour indices will only be shown if necessary.

Our goal is to derive the corrections to the SM couplings of quarks to $W^\pm$,
$Z$ and the neutral Higgs boson $H$ resulting from the mixing between the SM quarks
and heavy fermions after the latter have been integrated out {and subsequently the SM electroweak 
symmetry has been spontaneously broken.} To this end it
will be useful to decompose the relevant fundamental Lagrangian into pure
kinetic part, mass terms {for heavy fermions that are present before 
EWSB, Yukawa interaction and gauge interaction terms.} As
our main application will involve the RS model analysed in detail in
\cite{Albrecht:2009xr}, we will adopt the notation of this paper. Needless to
say the formulation below applies to any other model with the SM gauge group
at low energies, one Higgs doublet, ordinary quarks and leptons and heavy vector-like fermions as introduced above. 

\subsection{Kinetic Terms}
The kinetic terms for all the fermions in the theory are given as usual by
\bea\label{eq:kinetic-terms}
\mathcal{L}_\text{kin}&=&\bar\Psi_L(2/3)i\slash\partial\Psi_L(2/3)+\bar\Psi_R(2/3)i\slash\partial\Psi_R(2/3)
\nonumber\\
&+&\bar\Psi_L(-1/3)i\slash\partial\Psi_L(-1/3)+\bar\Psi_R(-1/3)i\slash\partial\Psi_R(-1/3)\,.
\eea
They are canonically normalised. However, as we will see in
Section~\ref{Sec:integrating out}, after the heavy fermions have been
integrated out 
{and EWSB took place,} the kinetic terms of ordinary quarks in (\ref{eq:kinetic-terms}) will acquire corrections and we will have to bring them back to the canonical form by properly redefining the ordinary quark fields.

\subsection{Mass Terms before EWSB}
The mass terms of the fermions {before EWSB} in our theory are given by
\be\label{eq:heavy mass}
\tilde{\mathcal{L}}_\text{mass}=-\bar\Psi_L(2/3)
\tilde{\mathcal{M}}(2/3)\Psi_R(2/3)-\bar\Psi_L(-1/3)
\tilde{\mathcal{M}}(-1/3)\Psi_R(-1/3)\,+\,h.c.\,.
\ee
Here $\tilde{\mathcal{M}}(2/3)$ and $\tilde{\mathcal{M}}(-1/3)$ are $3(N+1)\times
3(N+1)$ and $3(M+1)\times 3(M+1)$ {diagonal} matrices, respectively.
{The first three entries on the diagonal corresponding to SM quark masses vanish
at this stage, while the remaining entries are $\mathcal{O}(f)$ with $f$ being
the mass scale of heavy fermions.}
\subsection{Yukawa Interactions}
For our purposes it will be sufficient to write the Yukawa interactions 
showing explicitly only the lower component of the Higgs doublet that 
we denote by $\Phi$. Then
\bea
\mathcal{L}_\text{Y}&=&-\Phi\left[\bar\Psi_L(2/3)\mathcal{Y}(2/3)\Psi_R(2/3)\right.\nonumber\\
&+&\left.\bar\Psi_L(-1/3)\mathcal{Y}(-1/3)\Psi_R(-1/3)+\ \textrm{h.c.}\right]\,.
\label{eq:Yukawa}
\eea
Here ${\mathcal{Y}}(2/3)$ and ${\mathcal{Y}}(-1/3)$ are $3(N+1)\times
3(N+1)$ and $3(M+1)\times 3(M+1)$ complex matrices, respectively. 

\subsection{Neutral Currents}
The couplings of fermions to the linear combination of gauge fields that will be identified with the $Z$ boson after EWSB are described by the following neutral current
\bea
J_\mu(Z)&=&\bar\Psi_L(2/3)\gamma_\mu \mathcal{A}_L^{2/3}(Z)\Psi_L(2/3)\nonumber\\
&+&\bar\Psi_R(2/3)\gamma_\mu \mathcal{A}_R^{2/3}(Z)\Psi_R(2/3)\nonumber\\
&+&\bar\Psi_L(-1/3)\gamma_\mu \mathcal{A}_L^{-1/3}(Z)\Psi_L(-1/3)\nonumber\\
&+&\bar\Psi_R(-1/3)\gamma_\mu \mathcal{A}_R^{-1/3}(Z)\Psi_R(-1/3)\,.
\label{eq:J_mu_Z}
\eea
Here $\mathcal{A}_{L,R}^{2/3}(Z)$ and $\mathcal{A}_{L,R}^{-1/3}(Z)$ are again $3(N+1)\times3(N+1)$ and $3(M+1)\times3(M+1)$ matrices, respectively.

{In what follows it will be} useful to work with the building blocks
$\left[A_{L,R}(Z)\right]_{00}$, $\left[A_{L,R}(Z)\right]_{ij}$,
$\left[A_{L,R}(Z)\right]_{0j}$ and $\left[A_{L,R}(Z)\right]_{i0}$ which are
$3\times3$ matrices in flavour space. {Here the index $0$ denotes the
SM fermion, while the indices $i,j$ denote the heavy fermions.} 
{These matrices} have the following properties \cite{Arzt:1994gp}:
\bea\nonumber
i&\!\!)&\quad \left[A_{L,R}(Z)\right]_{00}\,\textrm{and}\ \left[A_{L,R}(Z)\right]_{ii}\ 
\textrm{are non-zero diagonal matrices}\label{eq:property-1}\\\nonumber
ii&\!\!)&\quad \left[A_{L,R}(Z)\right]_{ij}=0 \ \textrm{for}\ i\neq j\label{eq:property-2}\\\nonumber
iii&\!\!)&\quad \left[A_{L,R}(Z)\right]_{i0}=\left[A_{L,R}(Z)\right]_{0j}=0\,.\label{eq:property-3}
\eea
It should be emphasised that as opposed to \cite{Albrecht:2009xr} the coupling matrices given above are given still before electroweak symmetry breaking
and {some of their} entries that were non-vanishing in \cite{Albrecht:2009xr} because of the mixing of $Z$ with other gauge bosons are absent now. 

\subsection{Charged Currents}
The couplings of fermions to the linear combination of gauge fields that will be identified with the $W^\pm$ boson are described by the following charged current
\bea
J_\mu(W^+)&=&\bar\Psi_L(2/3)\gamma_\mu \mathcal{G}_L(W^+)\Psi_L(-1/3)\nonumber\\
&+&\bar\Psi_R(2/3)\gamma_\mu \mathcal{G}_R(W^+)\Psi_R(-1/3)\,,
\label{eq:J_mu_W}
\eea
where $\mathcal{G}_{L,R}(W^+)$ are this time $3(N+1)\times3(M+1)$ matrices. 
 Again, formula (\ref{eq:J_mu_W}) is written before EWSB and the properties {\it i}) - {\it iii}) given above are valid for the building blocks $\left[G_{L,R}(W^+)\right]_{00}$, $\left[G_{L,R}(W^+)\right]_{ij}$, $\left[G_{L,R}(W^+)\right]_{0j}$ and $\left[G_{L,R}(W^+)\right]_{i0}$.
In particular, we have $\left[G_{R}(W^+)\right]_{00}=0$ since in the SM $W^\pm$ does not couple to right-handed quarks.

\newsection{Integrating out Heavy Fermions and EWSB}\label{Sec:integrating out}
Having at hand all the relevant terms in the fundamental Lagrangian, 
we can construct the low-energy theory which involves only SM quark and
gauge boson fields and the Higgs field. There are several methods for 
achieving this goal. As in the present paper we are only interested in
corrections to the SM couplings we have found it to be most convenient to
integrate out the heavy fermions at tree level by using their EOM.
Inserting the solution for these equations in our fundamental Lagrangian
and expanding in powers of $1/f$ results in the effective Lagrangian of
which the $D=4$ part is the SM Lagrangian and the $D=6$ part is the one 
we are interested in. Performing then EWSB implies the replacement 
\be\label{eq:EWSB}
\Phi=\frac{1}{\sqrt{2}}\left[v+H\right],
\ee
where $H$ denotes the physical neutral Higgs and 
$v=246\gev$ is the vacuum expectation value of $\sqrt{2}\Phi$. Making this 
replacement  in the effective Lagrangian allows to find the corrections
to the SM couplings that result from the mixing with heavy vector-like
fermions.

As this procedure is well known~\cite{delAguila:2000aa,Burgess:1993vc}, we think that instead of presenting 
the details of this derivation it is more novel to find
a recipe for finding the corrections in question directly from our
fundamental Lagrangian of Section~2.

To this end we introduce
\begin{equation}\label{eq:mass-terms}
\mathcal{L}_\text{mass}=-\bar\Psi_L(2/3)\mathcal{M}(2/3)\Psi_R(2/3)-\bar\Psi_L(-1/3)\mathcal{M}(-1/3)\Psi_R(-1/3)\,+\,h.c.\,.
\end{equation}
Here $\mathcal{M}(2/3)$ and $\mathcal{M}(-1/3)$ are $3(N+1)\times
3(N+1)$ and $3(M+1)\times 3(M+1)$ matrices, respectively. They are
constructed by adding the $\tilde{\mathcal{L}}_\text{mass}$ in 
(\ref{eq:heavy mass})
and the mass terms resulting
from the Yukawa interactions in (\ref{eq:Yukawa}) after EWSB took place.

In what follows it will be useful to work with the building blocks of these matrices, $M_{00}(2/3)$, $M_{0i}(2/3)$, $M_{i0}(2/3)$, $M_{ij}(2/3)$ $(i,j=1,\dots ,N)$ and $M_{00}(-1/3)$, $M_{0i}(-1/3)$, $M_{i0}(-1/3)$, $M_{ij}(-1/3)$ $(i,j=1,\dots ,M)$, that are $3\times 3$ matrices in flavour space and where $M_{00}(2/3)$ and $M_{00}(-1/3)$ denote the mass matrices of ordinary quarks in the absence of the heavy fermions $U_{L,R}^i$ and $D_{L,R}^i$.

The matrices $\mathcal{M}(2/3)$ and $\mathcal{M}(-1/3)$ are complex and non-diagonal and have the following properties:
\begin{enumerate}
\item $M_{kk}=\mathcal{O}(f)$ where $f\gg v$ {is the mass scale of the heavy fermions}
\item $M_{00}=\mathcal{O}(v)$
\item $M_{ij}$ with $i\neq j$ are $\mathcal{O}(v)$ but could also vanish
\item $M_{0k}$ and $M_{k0}$ are generally $\mathcal{O}(v)$ but if $M_{0k}\neq 0$ then $M_{k0}=0$ and vice versa. This follows from the known property that only one of the chiralities of each vector-like fermion couples to the SM quarks through mass terms \cite{Arzt:1994gp}.
\end{enumerate}

In order to have a more transparent structure of resulting expressions we denote the $3\times3$ matrices with $j=k$ simply as
\be
M_{kk}(2/3)\equiv M_k(2/3)\,,\qquad M_{kk}(-1/3)\equiv M_k(-1/3)\,.
\ee
Then the solution to the EOM can be written before EWSB for the $-1/3$ charge heavy fields as follows

\begin{eqnarray}\label{eq:A1}
 D_L^k&=&-\left[M_k^{-1} \Phi Y_{0k}^\dagger-M_k^{-1}\Phi Y_{jk}^\dagger M_j^{-1}\Phi Y_{0j}^\dagger\right]d_L\,,\\\label{eq:A2}
 D_R^k&=&-\left[M_k^{-1} \Phi Y_{k0}-M_k^{-1}\Phi Y_{kj} M_j^{-1}\Phi Y_{j0}\right]d_R\,,
\end{eqnarray}

\noindent where we dropped on the r.h.s. terms that do not affect our final formulae. Here $Y_{ij}$ are $3\times 3$ submatrices of $\mathcal Y(-1/3)$ related to the submatrices $M_{ij}(-1/3)$ as follows 

\begin{equation}
 M_{ij}(-1/3)=\frac{v}{\sqrt{2}} Y_{ij}\,.
\end{equation}

\noindent Analogous formulae for the $+2/3$ charge heavy fields exist.

Having all these formulae at hand we are in the position to present a recipe
for finding corrections to the SM couplings at order $v^2/f^2$ directly from
the fundamental Lagrangian supplemented by the information contained 
in (\ref{eq:mass-terms}).
Concentrating then first on fermion-gauge boson couplings we just set $\Phi=\frac{v}{\sqrt 2}$. This results in the following steps of our recipe.

{\bf Step 1}: Express the heavy fields in terms of the light ones. 
For the $-1/3$ charge fields we have, using (\ref{eq:A1}) and (\ref{eq:A2})
\bea
D_L^k&=&-\left[M_k^{-1}M_{0k}^\dagger-M_k^{-1}M_{jk}^\dagger M_j^{-1}M_{0j}^\dagger \right]d_L\,,\label{eq:D_L}\\
D_R^k&=&-\left[M_k^{-1}M_{k0}-M_k^{-1}M_{kj}M_j^{-1}M_{j0}\right]d_R\,,\label{eq:D_R}
\eea
where the summation over $(k,j)$ indices with $k\neq j$ is understood and all mass matrices are for the down-quarks. The analogous formulae apply to up-quarks with $M_{ij}(-1/3)$ replaced by $M_{ij}(2/3)$. As we do not indicate in our notation that $M_{kj}$ are matrices and $D_L^k$ and $d_L$ are three dimensional vectors in flavour space, in order to avoid confusion let us just state with an example that $M_{jk}^\dagger$ stands for the hermitian conjugate of the $3\times3$ matrix $M_{jk}$.
The terms on the r.h.s.~of (\ref{eq:D_L}) and (\ref{eq:D_R}) are
$\mathcal{O}\left(v/f\right)$ and $\mathcal{O}\left(v^2/f^2\right)$. 

{\bf Step 2}: Redefine the SM fields to bring their kinetic terms
into canonical form. In the case of the SM down quark fields this 
amounts to the replacements
\bea
d_L&\to&\left(\mathbbm{1}-\frac{1}{2}M_{0k}M_k^{-2}M_{0k}^\dagger\right)d_L\,,\label{eq:d_L-shift}\\
d_R&\to&\left(\mathbbm{1}-\frac{1}{2}M_{k0}^\dagger M_k^{-2}M_{k0}\right)d_R\,,\label{eq:d_R-shift}
\eea
with analogous redefinitions for the up-quark fields.

Indeed inserting (\ref{eq:D_L}) and (\ref{eq:D_R}) into the kinetic terms of
the heavy fields in (\ref{eq:kinetic-terms}) we find that the light quark
kinetic terms are no longer canonically normalised. Keeping only the leading
$v^2/f^2$ terms, the canonical form of the kinetic terms is recovered after 
the transformations given in Step 2.
The fields on the r.h.s.~of (\ref{eq:d_L-shift}) and (\ref{eq:d_R-shift}) have now canonically normalised kinetic terms.

These two steps allow to derive directly the corrections to the SM fermion-gauge couplings. In order to derive the Higgs-fermion couplings, we have to generalise Step 1 by inserting the full expression for $\Phi$ in (\ref{eq:EWSB}) into (\ref{eq:A1}) and (\ref{eq:A2}). The results will be given in Section \ref{sec:nonderivative}.\\

This completes the integrating out of heavy fermions from the low energy theory which contains in addition to SM gauge fields and the neutral Higgs only the SM fermion fields.
What remains to be done is to insert the expressions (\ref{eq:D_L}) and (\ref{eq:D_R}), similar expressions involving $H$ and the corresponding expressions for the up-quarks into the fundamental Lagrangian supplemented by (\ref{eq:mass-terms}) and to perform the redefinitions of the fields as given by (\ref{eq:d_L-shift}) and (\ref{eq:d_R-shift}). The results are the $\mathcal O(v^2/f^2)$ corrections to the light fermion mass matrices, gauge couplings and Higgs couplings that we will summarise in the next section.

We would like to emphasise that the redefinitions of the light fields in
question are essential to obtain the correct effective couplings of quarks to
gauge bosons and the Higgs. In particular, omitting them would spoil the
custodial protection of various couplings as we will demonstrate below.

\boldmath
\newsection{Corrections to SM Mass Matrices and the $Z$, $W^\pm$, $H$ 
Couplings to SM Fermions}
\unboldmath
Proceeding along the steps outlined in the previous section it is
straightforward to calculate the impact of heavy fermions on the masses and
couplings of light fermions. Therefore we present only the final results. The
application to the RS model {considered in \cite{Albrecht:2009xr} will be performed in the 
following sections.}

\subsection{Mass Matrices}
For the mass matrices of down- and up-quarks, as defined in (\ref{eq:mass-terms}) but with heavy fields removed, we find the general expression $(k\neq j)$
\bea
M&=&M_{00}+M_{0k}M_k^{-1}M_{kj}M_j^{-1}M_{j0}\nonumber\\
&-&\frac{1}{2}\left[M_{0k}M_k^{-2}M_{0k}^\dagger M_{00}+M_{00}M_{k0}^\dagger M_k^{-2}M_{k0}\right]\,,
\label{eq:mass-matrix}
\eea 
with the first correction originating from the pure heavy mass terms and the
{second correction containing $M_{00}$} from the redefinitions of the light quark fields. Here and in the following formulae (\ref{eq:A_L_Z})-(\ref{eq:G_R_W}), (\ref{eq:Y_H})-(\ref{eq:correction_H}) and (\ref{eq:Y_non-diag}) summation over repeated indices is understood.

\subsection{Couplings to Gauge Bosons}
For the couplings to neutral gauge bosons as defined in (\ref{eq:J_mu_Z}) but with heavy fermions removed we find
\bea
A_L(Z)&=&\left[A_L(Z)\right]_{00}+M_{0k}M_k^{-1}\left[A_L(Z)\right]_{kk}M_k^{-1}M_{0k}^\dagger\nonumber\\
&-&\frac{1}{2}M_{0k}M_k^{-2}M_{0k}^\dagger\left[A_L(Z)\right]_{00}\nonumber\\
&-&\frac{1}{2}\left[A_L(Z)\right]_{00}M_{0k}M_k^{-2}M_{0k}^\dagger\,,
\label{eq:A_L_Z}
\eea
\bea
A_R(Z)&=&\left[A_R(Z)\right]_{00}+M_{k0}^\dagger M_k^{-1}\left[A_R(Z)\right]_{kk}M_k^{-1}M_{k0}\nonumber\\
&-&\frac{1}{2}M_{k0}^\dagger M_k^{-2}M_{k0}\left[A_R(Z)\right]_{00}\nonumber\\
&-&\frac{1}{2}\left[A_R(Z)\right]_{00}M_{k0}^\dagger M_k^{-2}M_{k0}\,.
\label{eq:A_R_Z}
\eea
These formulae apply to both charge $+2/3$ and $-1/3$ quarks with appropriate use of $\left[A_{L,R}^{2/3}(Z)\right]_{\alpha\beta}$ or $\left[A_{L,R}^{-1/3}(Z)\right]_{\alpha\beta}$ couplings, respectively, and similarly for the mass matrices, where $(\alpha,\beta=0,i)$.\\

For the couplings to charged gauge bosons as defined in (\ref{eq:J_mu_W}) we find
\begin{eqnarray}\label{eq:G_L_W}
G_L(W^+)&=&\left[G_L(W^+)\right]_{00}+M_{0k}(2/3)M_k^{-1}(2/3)\left[G_L(W^+)\right]_{kk}M_k^{-1}(-1/3)M_{0k}^\dagger(-1/3)\nonumber\\
&-&\frac{1}{2}M_{0k}(2/3)M_k^{-2}(2/3)M_{0k}^\dagger(2/3)\left[G_L(W^+)\right]_{00}\nonumber\\
&-&\frac{1}{2}\left[G_L(W^+)\right]_{00}M_{0k}(-1/3)M_k^{-2}(-1/3)M_{0k}^\dagger(-1/3)\,,\\\label{eq:G_R_W}
G_R(W^+)&=&M_{k0}^\dagger(2/3)
M_k^{-1}(2/3)\left[G_R(W^+)\right]_{kk}M_k^{-1}(-1/3)M_{k0}(-1/3)\,.
\end{eqnarray}

We note that the equations (\ref{eq:A_L_Z})-(\ref{eq:G_L_W}) have the same structure. The 
first corrections on the r.h.s.~originate in the interactions of the heavy fermion fields
with the SM gauge bosons and the remaining terms in these equations are the consequence of
the redefinitions of the light fields given in (\ref{eq:d_L-shift}) and (\ref{eq:d_R-shift}).

An exception is the coupling $G_R(W^+)$ which vanishes at the leading order so
that the redefinitions of the light fields do not matter at order 
$v^2/f^2$.

\subsection{Couplings to the Higgs Boson}\label{sec:nonderivative}
The fermion-Higgs couplings can be summarised by
\begin{equation}
 \mathcal L_Y=-\frac{H}{\sqrt 2}\left[\bar u_L Y(2/3) u_R+\bar d_L Y(-1/3) d_R+\ \textrm{h.c.}\right]\,,
\end{equation}
with 
\be
Y=Y_1+Y_2\,.\label{eq:Y_1_2}
\ee
Here $Y_1$ is obtained directly from terms involving no derivatives, while $Y_2$ results from terms involving derivatives $\partial_\mu H$, $\slash\partial u_{L,R}$ and $\slash\partial d_{L,R}$.

We begin with $Y_1$.
Using the notation $Y_{00}$, $Y_{0j}$, $Y_{j0}$, $Y_{ij}$ for the $3\times3$ sub-matrices of $\mathcal Y(2/3)$ and $\mathcal Y(-1/3)$ in
(\ref{eq:Yukawa}) we find for both up- and down-quark Yukawa couplings ($Y(2/3)$ and $Y(-1/3)$ respectively) the common expression
\bea
Y_1&=&Y_{00}+M_{0k}M_k^{-1}Y_{kj}M_j^{-1}M_{j0}\nonumber\\
&-&\frac{1}{2}\left[M_{0k}M_k^{-2}M_{0k}^\dagger Y_{00}+Y_{00}M_{k0}^\dagger M_k^{-2}M_{k0}\right]\nonumber\\
&+&Y_{0k}M_k^{-1}M_{kj}M_j^{-1}M_{j0}+M_{0j}M_j^{-1}M_{jk}M_k^{-1}Y_{k0}\,,
\label{eq:Y_H}
\eea
where for the up-quarks $Y_{00}(2/3)$, $M_{ij}(2/3)$ etc.~should be inserted on the r.h.s.~of this formula,
and $Y_{00}(-1/3)$, $M_{ij}(-1/3)$ etc.~for the down-quarks, respectively.

Next, working with the expressions (\ref{eq:A1}), (\ref{eq:A2})  
that are given before EWSB 
and the corresponding expressions for the up-quarks 
we find that there are also terms involving derivatives of either the light quark fields or the Higgs boson contributing to the Higgs coupling. For these couplings involving $\partial_\mu q_{L,R}$ and $\partial_\mu H$ we obtain ($q=u,d$)
\begin{eqnarray}
\mathcal{L}
_\text{der.}&=&\frac{1}{\sqrt{2}}(i\partial_\mu H)\left[\bar q_L\gamma^\mu M_{0k}M_k^{-2}Y_{0k}^\dagger q_L+\bar q_R \gamma^\mu M_{k0}^\dagger M_k^{-2}Y_{k0}q_R\right]\nonumber\\
&+&\sqrt{2}H\left[\bar q_LM_{0k}M_k^{-2}Y_{0k}^\dagger (i\slash\partial q_L)+\bar q_RM_{k0}^\dagger M_k^{-2}Y_{k0}(i\slash\partial q_R)\right]\,,\label{eq:derivative_Higgs_couplings}
\end{eqnarray}
where for the up-quarks $M_{00}(2/3)$, $M_{ij}(2/3)$ etc.~should be inserted on the r.h.s.~of this formula, and $M_{00}(-1/3)$, $M_{ij}(-1/3)$ etc.~for the down-quarks, respectively.

Integrating by parts the terms in the first line of (\ref{eq:derivative_Higgs_couplings}) and employing the EOM for the SM quarks then yields
\begin{eqnarray}
\mathcal{L}
_\text{der.}=\frac{1}{\sqrt{2}}H&&\hspace{-12pt}\left[\bar q_RM_{00}^\dagger M_{0k}M_k^{-2}Y_{0k}^\dagger q_L+\bar q_LM_{0k}M_k^{-2}Y_{0k}^\dagger M_{00}q_R\right.\nonumber\\
&+&\left.\bar q_LM_{00} M_{k0}^\dagger M_k^{-2}Y_{k0}q_R+\bar q_RM_{k0}^\dagger M_k^{-2}Y_{k0}M_{00}^\dagger q_L\right]\,,\label{eq:der_H_2}
\end{eqnarray}
implying
\begin{equation}
Y_2=-M_{0k}M_k^{-2}M_{0k}^\dagger Y_{00}-Y_{00} M_{k0}^\dagger M_k^{-2}M_{k0}\,.\label{eq:correction_H}
\end{equation}

\subsection{Going to the Quark Mass Eigenstate Basis}
The formulae (\ref{eq:mass-matrix})-(\ref{eq:G_R_W}), (\ref{eq:Y_H}) ad (\ref{eq:correction_H}) are still given for quarks in the flavour basis. In order to find the corresponding formulae for the quark mass eigenstates we have to diagonalise the mass matrices $M(Q)$, ($Q=-1/3,\,2/3$) in (\ref{eq:mass-matrix}),
\bea
M_\text{diag}(-1/3)&=&\mathcal{D}_L^\dagger\,M(-1/3)\,\mathcal{D}_R\,,\label{eq:M_diag_down}\\
M_\text{diag}(2/3)&=&\mathcal{U}_L^\dagger\,M(2/3)\,\mathcal{U}_R\,.\label{eq:M_diag_up}
\eea
Then the mass eigenstates are given by
\bea
\left(d_{L,R}\right)_\text{mass}&=&\mathcal{D}_{L,R}^\dagger\, d_{L,R}\,,\label{eq:mass_eigenstate_down}\\
\left(u_{L,R}\right)_\text{mass}&=&\mathcal{U}_{L,R}^\dagger\, u_{L,R}\,.\label{eq:mass_eigenstate_up}
\eea
In the mass eigenstate basis the neutral 
couplings read
\bea
\left[A_{L,R}^{-1/3}(Z)\right]_\text{mass}&=&\mathcal{D}_{L,R}^\dagger\,A_{L,R}^{-1/3}(Z)\,\mathcal{D}_{L,R}\,,\label{eq:A_Z_down_mass_eigenstates}\\
\left[A_{L,R}^{2/3}(Z)\right]_\text{mass}&=&\mathcal{U}_{L,R}^\dagger\,A_{L,R}^{2/3}(Z)\,\mathcal{U}_{L,R}\,.\label{eq:A_Z_up_mass_eigenstates}
\eea
For the charged couplings we find
\be
\left[G_{L,R}(W^+)\right]_\text{mass}=\mathcal{U}_{L,R}^\dagger\,G_{L,R}(W^+)\,\mathcal{D}_{L,R}\,.\label{eq:G_W_mass_eigenstates}
\ee
For the Higgs couplings we have in analogy to (\ref{eq:M_diag_down}) and (\ref{eq:M_diag_up})
\bea
Y_\text{mass}^{-1/3}&=&\mathcal{D}_{L}^\dagger\,Y^{-1/3}\, \mathcal{D}_{R}\,,\label{eq:Y_H_down_mass_eigenstates}\\
Y_\text{mass}^{2/3}&=&\mathcal{U}_{L}^\dagger\,Y^{2/3}\, \mathcal{U}_{R}\,.\label{eq:Y_H_up_mass_eigenstates}
\eea

The formulae (\ref{eq:A_Z_down_mass_eigenstates})-(\ref{eq:Y_H_up_mass_eigenstates}) together with (\ref{eq:A_L_Z})-(\ref{eq:G_R_W}), (\ref{eq:Y_H}) and (\ref{eq:correction_H}) are the main model independent results of the present paper. They summarise the couplings of SM quarks to $W^\pm$, $Z$ and the Higgs after the inclusion of $\mathcal O(v^2/f^2)$ corrections from the mixing of these quarks with heavy vector-like fermions. It should be emphasised that these formulae are valid for both flavour diagonal and non-diagonal couplings of $Z$ and $H$ to SM quarks. However, the formulae for the non-diagonal Higgs couplings simplify in the mass eigenstate basis as the first four terms on the r.h.s.~of (\ref{eq:Y_H}) have the same flavour structure as $M$ in (\ref{eq:mass-matrix}) and consequently are diagonalised when going to the mass eigenstate basis. Combining (\ref{eq:Y_H}) and (\ref{eq:correction_H}) we consequently find
\bea
\left[Y_\text{mass}^{-1/3}\right]_\text{non-diag}&=&\mathcal{D}_L^\dagger\left[Y_{0k}M_k^{-1}M_{kj}M_j^{-1}M_{j0}+M_{0j}M_j^{-1}M_{jk}M_k^{-1}Y_{k0}\right.\nonumber\\
&-&\left.M_{0k}M_k^{-2}M_{0k}^\dagger Y_{00}-Y_{00} M_{k0}^\dagger M_k^{-2}M_{k0}\right]\mathcal{D}_R\nonumber\\
&=&-2\mathcal{D}_L^\dagger Y_{00}\mathcal{D}_R\,,\label{eq:Y_non-diag}
\eea
and similarly for the up-quarks.

{As the diagonalising matrices $\mathcal D_{L,R}$ differ from the ones that diagonalise $Y_{00}$ flavour changing neutral Higgs-fermion interactions are generated at  $\mathcal O(v^2/f^2)$.}

\newsection{Custodial Protection of Neutral Couplings in Explicit Terms}
\subsection{Preliminaries\label{sec:Custodial_protection_preliminaries}}
Before {we apply} our formalism to the RS model with the fermion
representations of \cite{Albrecht:2009xr}, it will be useful to recall first
the basis of the custodial protection of $Z$-couplings {developed in}
\cite{Agashe:2006at}, and its generalisation to flavour violating left-handed down-quark
couplings of $Z$ demonstrated in
\cite{Blanke:2008zb,Blanke:2008yr}. In this context we will point out that
this kind of protection is also valid for the flavour-conserving and
flavour-violating right-handed up-quark couplings.

The couplings of a given fermion $F$ to the $Z$ boson are protected if this fermion is an eigenstate of $P_{LR}$. This implies the following condition \cite{Agashe:2006at} for the quantum numbers of $F$ under the gauge group (\ref{eq:gaugegroup})

\begin{equation}\label{eq:protection}
T_L=T_R\,,\qquad T_L^3=T_R^3\qquad (P_{LR})\,.
\end{equation}
The authors of \cite{Agashe:2006at} have also noticed that for a fermion with $T_L\neq T_R$ satisfying 

\begin{equation}\label{eq:protectionPC}
T_L^3=T_R^3=0\qquad(P_C)
\end{equation}
its $Z$-couplings are protected as well. This time a discrete subgroup of the custodial $SU(2)_V$, denoted by $P_C$, is responsible for this protection. In Table \ref{tab:fieldcontentSM} we recall the quantum numbers $Q$, $T_{L,R}$ and $T_{L,R}^3$ of the SM quarks. {In Table \ref{tab:fieldcontent} these quantum numbers for the left-handed KK-fermions in the model of \cite{Albrecht:2009xr} are summarised}. The right-handed KK-fermions have the same quantum numbers except {for reversed BCs}.

In \cite{Agashe:2006at} only the $Zb\bar b$ coupling has been discussed. However, when {the fermions of each flavour are put into the same representations}, the $P_{LR}$ and $P_C$ symmetries are also active for flavour violating couplings of $Z$. This has been pointed out in \cite{Blanke:2008zb,Blanke:2008yr} for left-handed couplings of down-quarks relevant for the phenomenology in \cite{Blanke:2008zb,Blanke:2008yr}. Here we emphasise that also some right-handed couplings can be protected in this manner. 

Indeed, the inspection of Tables \ref{tab:fieldcontentSM} and \ref{tab:fieldcontent} in conjunction with (\ref{eq:protection}) and (\ref{eq:protectionPC}) reveals the protection of the following $Z$ couplings:

\begin{enumerate}\label{protection:i}
\item left-handed couplings of SM down-quarks
\item right-handed couplings of SM up-quarks\label{protection:iii}
\item couplings of $\chi_{L,R}^{u^i}$\label{protection:iv}
\item couplings of $U_{L,R}^{\prime i}$ and of $U_{L,R}^{\prime\prime i}$\,.
\end{enumerate}
It should be noted that whereas the {protection in 1.-3.~follows} from (\ref{eq:protection}), the last protection is guaranteed by (\ref{eq:protectionPC}). 

In what follows we would like to {inspect} this protection in explicit terms by analysing flavour violating couplings of $Z$ generated at $\mathcal{O}(v^2/M_\text{KK}^2)$ through the mixing of $Z$ with neutral KK-gauge bosons. While for the left-handed SM couplings this has already been demonstrated in \cite{Blanke:2008zb,Blanke:2008yr}, the {analogous} presentation for right-handed SM quark couplings and the couplings of KK-fermions is new.

To this end and also for our discussion of the effects of KK-fermions on the SM gauge couplings it will be useful to recall two charge factors {for a fermion $F$ with isospins $T_L^3$ and $T_R^3$ and electric charge $Q$} introduced in \cite{Albrecht:2009xr}:
\begin{eqnarray}\label{eq:g}
g^{4D}_Z(F)&=&\frac{g^{4D}}{\cos{\psi}}\left[T_L^3-(\sin\psi)^2 Q\right]\,,\\\label{eq:kappa}
\kappa^{4D}(F)&=&\frac{g^{4D}}{\cos\phi}\left[T_R^3-(Q-T_L^3)\sin^2\phi\right]\,,
\end{eqnarray}
{where $g_Z^{4D}(F)$ and $\kappa^{4D}(F)$ denote the couplings of $F$ to the $Z^{(0),(1)}$ and $Z_X^{(1)}$ gauge bosons, respectively.}
{The angles} $\psi$ and $\phi$ are related by 
\be\label{eq:twoangles}
\cos\psi=\frac{1}{\sqrt{1+\sin^2\phi}}\,,\qquad\sin\psi=\frac{\sin\phi}{\sqrt{1+\sin^2\phi}}\,.
\ee

Starting from these general equations it is straightforward to compute the charge factors for the down quarks, using the quantum numbers given in Tables~\ref{tab:fieldcontentSM} and \ref{tab:fieldcontent}. The results for $g_{Z,L}^{4D}(d)$, $g_{Z,R}^{4D}(d)$,
$\kappa_1^{4D}(d)$ and $\kappa_5^{4D}(d)$ have been collected in the Appendix,
{where the charge factors for the up quarks $g_{Z,L}^{4D}(u)$,
  $g_{Z,R}^{4D}(u)$, $\kappa_1^{4D}(u)$ and $\kappa_3^{4D}(u)$ can also be 
found.}

As we will see below the ``magic'' formula
\begin{equation}\label{eq:combination}
g^{4D}_Z(F)-\cos\phi\cos\psi\kappa^{4D}(F)=0\,
\end{equation}
summarises compactly all protections discussed in this section. Indeed, using (\ref{eq:g}), (\ref{eq:kappa}) and Tables \ref{tab:fieldcontentSM} and \ref{tab:fieldcontent}, we verify that the protected couplings listed in 1.-4. do satisfy (\ref{eq:combination}). 

In the {remainder} of this section we will demonstrate that in the approximation of neglecting the violation of $P_{LR}$ and $P_C$ symmetries through boundary conditions on the UV brane, all protected couplings are proportional to the ``magic'' combination on the l.h.s.~of (\ref{eq:combination}).
\begin{table}
\begin{center}
\begin{tabular}{|c|r|r|r|}
\hline
Field                 &     Charge $Q$             &         Isospin $T_L^3$    &    Isospin $T_R^3$\\\hline
$q^{u_i(0)}_L(++)$       &     $\frac{2}{3}$          &    $\frac{1}{2}$           &    $-\frac{1}{2}$\\\hline
$q^{d_i(0)}_L(++)$       &    $-\frac{1}{3}$          &      $-\frac{1}{2}$        &      $-\frac{1}{2}$\\\hline
$u^i_R(++)$           &     $\frac{2}{3}$          &       $0$                  &         $0$\\\hline
$D^{i}_R(++)$         &       $-\frac{1}{3}$       &      $0$                   &            $-1$\\\hline
\end{tabular}
\renewcommand{\arraystretch}{1.0}
\caption{SM  quark content of the theory.}\label{tab:fieldcontentSM}
\end{center}
\end{table}
\begin{table}
\begin{center}
\begin{tabular}{|c|r|r|r|}
\hline
Field                 &     Charge $Q$             &         Isospin $T_L^3$    &    Isospin $T_R^3$\\\hline
$\chi^{u_i}_L(-+)$    &    $\frac{5}{3}$           &   $\frac{1}{2}$            &     $\frac{1}{2}$\\\hline
$\chi^{d_i}_L(-+)$    &    $\frac{2}{3}$           &    $-\frac{1}{2}$          &    $\frac{1}{2}$\\\hline
$q^{u_i}_L(++)$       &     $\frac{2}{3}$          &    $\frac{1}{2}$           &    $-\frac{1}{2}$\\\hline
$q^{d_i}_L(++)$       &    $-\frac{1}{3}$          &      $-\frac{1}{2}$        &      $-\frac{1}{2}$\\\hline
$u^i_L(--)$           &     $\frac{2}{3}$          &       $0$                  &         $0$\\\hline
$\psi^{\prime i}_L(+-)$&       $\frac{5}{3}$       &      $1$                   &         $0$\\\hline
$\psi^{\prime\prime i}_L(+-)$&$\frac{5}{3}$        &      $0$                   &          $1$\\\hline
$U^{\prime i}_L(+-)$  &   $\frac{2}{3}$            &       $0$                  &           $0$\\\hline
$U^{\prime\prime i}_L(+-)$&$\frac{2}{3}$           &       $0$                  &           $0$\\\hline
$D^{\prime i}_L(+-)$  &     $-\frac{1}{3}$         &       $-1$                 &            $0$\\\hline
$D^{i}_L(--)$         &       $-\frac{1}{3}$       &      $0$                   &            $-1$\\\hline
\end{tabular}
\renewcommand{\arraystretch}{1.0}
\caption{Heavy  quark  content of the theory.  We show only 
 the left-handed quarks as the quantum numbers of the the right-handed heavy 
quarks
are the same and only their parities on the boundaries have to be reversed.
\label{tab:fieldcontent}}
\end{center}
\end{table}

\boldmath
\subsection{The Flavour Non-Diagonal $Zd_{L,R}^i\bar d_{L,R}^j$ Couplings}
\label{sec:non-diag_Z_d}
\unboldmath
The couplings of $Z$ to down-quarks in the absence of mixing with KK fermions are given up to an irrelevant factor $-i$ by \cite{Albrecht:2009xr}
\be
\Delta_{L,R}^{ij}(Z)=\frac{M_Z^2}{M_\text{KK}^2}\left[-\mathcal I_1^+\Delta_{L,R}^{ij}(Z^{(1)})+\mathcal I_1^-\cos\phi\cos\psi\Delta_{L,R}^{ij}(Z_X^{(1)})\right]\,.
\label{eq:Delta_L_R}
\ee
All the {ingredients of this equation} are defined in Appendix~A of~\cite{Blanke:2008yr}. We only recall that
\be
\mathcal I_1^+=\mathcal I_1^-
\label{eq:I_1}
\ee
up to the different boundary conditions of the shape functions of the gauge eigenstates $Z^{(1)}$ and $Z_X^{(1)}$ on the UV brane. Moreover, $\Delta_{L,R}^{ij}(Z^{(1)})$ and $\Delta_{L,R}^{ij}(Z_X^{(1)})$ are the elements of the $3\times3$ coupling matrices
\be
\hat\Delta_{L,R}(V)=\mathcal{D}_{L,R}^\dagger\hat\varepsilon_{L,R}(V)\mathcal{D}_{L,R}\qquad\left(V=Z^{(1)},Z_X^{(1)}\right)\,,
\ee
with $\mathcal{D}_L$ and $\mathcal{D}_R$ being the left- and right-handed down-type flavour mixing matrices that are used to diagonalise the down-quark mass matrices as already stated in (\ref{eq:M_diag_down}). $\hat\varepsilon_{L,R}(V)$ are diagonal {coupling} matrices. The diagonal $3\times3$ matrices $\hat\varepsilon_{L,R}(V)$ are given, up to again different boundary conditions for the shape functions of $Z^{(1)}$ and $Z_X^{(1)}$ on the UV brane, by a universal diagonal $3\times3$ matrix multiplied by a flavour independent charge factor that distinguishes $L$ from $R$ and $Z^{(1)}$ from $Z_X^{(1)}$ couplings. Thus in the approximation of neglecting the difference {in} the boundary conditions, as already done in (\ref{eq:I_1}), we can write\footnote{Note that this approximation is only valid for off-diagonal terms in $\hat\Delta_{L,R}(Z)$.}
\bea
\Delta_L^{ij}(Z)&=&F_L^{ij}(Z)\left[-g_{Z,L}^{4D}(d)+\cos\phi\cos\psi\kappa_1^{4D}(d)\right]\,,\label{eq:Delta_L_Z}\\
\Delta_R^{ij}(Z)&=&F_R^{ij}(Z)\left[-g_{Z,R}^{4D}(d)+\cos\phi\cos\psi\kappa_5^{4D}(d)\right]\,.\label{eq:Delta_R_Z}
\eea
The functions $F_{L,R}^{ij}$ are given by 
\bea
F_L^{ij}(Z)&=&\frac{M_Z^2}{M_\text{KK}^2}\frac{{\cal I}_1^+}{L}\left(\mathcal{D}_L\right)_{ki}^\ast\left(\mathcal{D}_L\right)_{kj}\int e^{ky}\left[f_L^{(0)}(y,c_1^k)\right]^2g^{}(y)dy\,,\\
F_R^{ij}(Z)&=&\frac{M_Z^2}{M_\text{KK}^2}\frac{{\cal I}_1^+}{L}\left(\mathcal{D}_R\right)_{ki}^\ast\left(\mathcal{D}_R\right)_{kj}\int e^{ky}\left[f_R^{(0)}(y,c_3^k)\right]^2g^{}(y)dy\,,
\eea
{where $g^{}(y)$ denotes the shape function of the first KK-excitation of a gauge boson with (++) boundary condition, and $f_L^{(0)}$, $f_R^{(0)}$ are the shape functions of the SM down-quarks.}

It is evident {from the couplings and charge factors given in the Appendix} and from (\ref{eq:twoangles}) that we have
\be
g_{Z,L}^{4D}(d)-\cos\phi\cos\psi\kappa_1^{4D}(d)=0\,,
\label{eq:coupling_constant_cancellation}
\ee
which signals the protection of the $Zd_L^i\bar d_L^j$ couplings. On the other hand, the $Zd_R^i\bar d_R^j$ couplings are not protected as the two terms in parentheses in (\ref{eq:Delta_R_Z}) do not cancel each other.

\boldmath
\subsection{The Flavour Non-Diagonal $Zu_{L,R}^i\bar u_{L,R}^j$ Couplings}
\label{sec:non-diag_Z_u}
\unboldmath
The couplings $Zu_{L,R}^i\bar u_{L,R}^j$ have the same structure as $Zd_{L,R}^i\bar d_{L,R}^j$ and differ from them only through new overlap integrals, resulting in $H_{L,R}^{ij}(Z)$ in place of $F_{L,R}^{ij}(Z)$ in (\ref{eq:Delta_L_Z}) and (\ref{eq:Delta_R_Z}), and charge factors. Consequently we find
\bea
\tilde\Delta_L^{ij}(Z)&=&H_L^{ij}(Z)\left[-g_{Z,L}^{4D}(u)+\cos\phi\cos\psi\kappa_1^{4D}(u)\right]\,,\label{eq:Delta_tilde_L_Z}\\
\tilde\Delta_R^{ij}(Z)&=&H_R^{ij}(Z)\left[-g_{Z,R}^{4D}(u)+\cos\phi\cos\psi\kappa_3^{4D}(u)\right]\,,\label{eq:Delta_tilde_R_Z}
\eea
where $H_{L,R}^{ij}(Z)$ are given as follows:
\bea
H_L^{ij}(Z)&=&\frac{M_Z^2}{M_\text{KK}^2}\frac{{\cal I}_1^+}{L}\left(\mathcal{U}_L\right)_{ki}^\ast\left(\mathcal{U}_L\right)_{kj}\int e^{ky}\left[f_L^{(0)}(y,c_1^k)\right]^2g^{}(y)dy\,,\\
H_R^{ij}(Z)&=&\frac{M_Z^2}{M_\text{KK}^2}\frac{{\cal I}_1^+}{L}\left(\mathcal{U}_R\right)_{ki}^\ast\left(\mathcal{U}_R\right)_{kj}\int e^{ky}\left[f_R^{(0)}(y,c_2^k)\right]^2g^{}(y)dy\,.
\eea

Using the general expressions for the charge factors (\ref{eq:g}), (\ref{eq:kappa}) and the quantum numbers for the up-quarks in 
Table~\ref{tab:fieldcontentSM} (or simply the explicit formulae given in the Appendix), we find
 \be
g_{Z,R}^{4D}(u)-\cos\phi\cos\psi\kappa_3^{4D}(u)=0\,,
\label{eq:coupling_constant_cancellation_2}
\ee
which signals the protection also of the $Zu_R^i\bar u_R^j$ couplings. On the other hand, the $Zu_L^i\bar u_L^j$ couplings are not protected as the two terms in parentheses in (\ref{eq:Delta_tilde_L_Z}) do not cancel each other. We will return to both couplings in Section~\ref{sec:Numerics}.

Proceeding in the same manner one can show that also the (vector-like) flavour
diagonal and off-diagonal couplings of $Z$ to the charge $5/3$ quarks
$\chi_{L,R}^{u^i}$ and to the charge $2/3$ quarks $U_{L,R}^{\prime i}$ and
$U_{L,R}^{\prime\prime i}$ {(see Table~\ref{tab:fieldcontent})} 
are protected by the custodial symmetry. Indeed the coupling of these three
fields to the $Z$ boson is proportional again to the ``magic'' combination in
(\ref{eq:combination}) which vanishes for all the three fields as we have
already 
mentioned above. {The charge factors of these fields are given explicitly in the Appendix}.

\boldmath
\subsection{Couplings of $Z_H$ and $Z^\prime$ to SM Quarks}\label{sec:ZHZprimeSMquarks}
\unboldmath
It will also be of interest to have a brief look at $Z_H$ and $Z^\prime$ couplings to SM quarks.
{Neglecting the mixing with KK fermions,} the couplings of $Z_H$ and $Z^\prime$ to down-quarks are given by
\bea
\hat\Delta_{L,R}(Z_H)&=&\cos\xi\hat\Delta_{L,R}(Z^{(1)})+\sin\xi\hat\Delta_{L,R}(Z_X^{(1)})\label{eq:Delta_Z_H}\,,\\
\hat\Delta_{L,R}(Z^\prime)&=&-\sin\xi\hat\Delta_{L,R}(Z^{(1)})+\cos\xi\hat\Delta_{L,R}(Z_X^{(1)})\,,
\label{eq:Delta_Z_prime}
\eea
with $\cos\xi$ and $\sin\xi$ given explicitly in \cite{Albrecht:2009xr}. In the limit of exact $P_{LR}$ symmetry one has
\be\label{eq:anglerelation}
\frac{\cos\xi}{\sin\xi}=\cos\phi\cos\psi\,,
\ee
and with (\ref{eq:I_1}), the formula (\ref{eq:Delta_Z_prime}) for $\hat\Delta_{L}^{ij}(Z^\prime)$ reduces to $\Delta_L^{ij}(Z)$ in (\ref{eq:Delta_L_R}) up to an overall factor. Consequently, both flavour diagonal and non-diagonal $Z^\prime$ couplings to left-handed down-quarks are protected. Analogous considerations using the relevant couplings given in the Appendix show that the right-handed couplings of down-quarks are not protected by the $P_{LR}$ symmetry. This is also the case for all the $Z_H$ couplings to down-quarks.

If we perform exactly the same type of analysis for the couplings of $Z_H$ and $Z^\prime$ to the SM up-quarks, we discover that\footnote{Here the same notation as in~\cite{Albrecht:2009xr} is adopted.}

\begin{eqnarray}
\hat{\tilde\Delta}_{R}(Z^\prime)&\sim&-g_{Z,R}^{4D}(u)\sin\xi\,\underset{00}{\mathcal{R}_2^i}(++)_R+\kappa_{3}^{4D}(u)\cos\xi\,\underset{00}{\mathcal{P}_2^i}(++)_R\label{eq:Delta_Z_prime up_right}\,,\\
\hat{\tilde\Delta}_{L}(Z^\prime)&\sim&-g_{Z,L}^{4D}(u)\sin\xi\,\underset{00}{\mathcal{R}_1^i}(++)_L+\kappa_{1}^{4D}(u)\cos\xi\,\underset{00}{\mathcal{P}_1^i}(++)_L\label{eq:Delta_Z_prime up_left}\,,
\end{eqnarray}
with $\underset{nm}{\mathcal{R}_k^i}(++)_{L,R}$ and $\underset{nm}{\mathcal{P}_k^i}(++)_{L,R}$ given by
\begin{eqnarray}\label{eq:intR}
\underset{nm}{\mathcal{R}_k^i}(BC)_{L,R} &=& \frac{1}{L} \int_0^L dy \, e^{ky}  f_{L,R}^{(n)}(y,c_k^i,BC)
 f_{L,R}^{(m)}(y,c_k^i,BC) \, g(y)\,,\\
\underset{nm}{\mathcal{P}_k^i}(BC)_{L,R}  &=& \frac{1}{L} \int_0^L dy \, e^{ky}  f_{L,R}^{(n)}(y,c_k^i,BC)
 f_{L,R}^{(m)}(y,c_k^i,BC) \,\tilde g(y)\,,\label{overlapR3}
\end{eqnarray}
where $\tilde g(y)$ and $g(y)$ are the shape functions of the gauge bosons with $(-+)$ and $(++)$ BC, respectively, and $f_{L,R}^{(n)}$ of the KK-fermions.

The corresponding couplings to $Z_H$ are simply obtained from (\ref{eq:Delta_Z_prime up_right}) and (\ref{eq:Delta_Z_prime up_left}) with the prescription $\cos\xi\rightarrow \sin\xi$, $\sin\xi\rightarrow -\cos\xi$~\cite{Albrecht:2009xr}.

In the limit of exact $P_{LR}$ symmetry the relation (\ref{eq:anglerelation}) holds, in addition to 
\begin{equation}
\underset{00}{\mathcal{R}_k^i}(++)_{L,R}=\underset{00}{\mathcal{P}_k^i}(++)_{L,R}\,.
\end{equation}

Using the charge factors of the up-quarks given in the Appendix, it is straightforward to see that also the diagonal and non-diagonal couplings of $Z^\prime$ to right-handed up-quarks are protected. Analogous considerations show that the left-handed couplings of up-quarks are not protected. The same holds for all the couplings of $Z_H$ to up-quarks.

{Based on} analogous considerations we can conclude that also the couplings of $Z^\prime$ with $U^\prime_{L,R}$, $U^{\prime\prime}_{L,R}$ and $\chi_{L,R}^{u^i}$ are protected by the custodial symmetry.

\newsection{Impact of KK Fermions on Neutral Couplings\label{sec:Impact_neutral}}
\subsection{Preliminaries}
In this section we will analyse the impact of KK fermions on the couplings of $Z$, demonstrating explicitly that the protection of $Zd_L^i\bar d_L^j$ and $Zu_R^i\bar u_R^j$ is maintained even in the presence of KK contributions. This is to be expected, as the fermion representations are symmetric under $P_{LR}$, {and above all $d_L$ and $u_R$ are $P_{LR}$-eigenstates}. Still it is instructive to {inspect} this protection in explicit terms.
We will also calculate the impact of KK fermions on the right-handed couplings $Zd_R^i\bar d_R^j$ and also on $Zu_L^i\bar u_L^j$. As these couplings are not protected by the $P_{LR}$ symmetry, also the effect of KK fermions, in particular in the case of the top-quark couplings, will be significant.

\boldmath
\subsection{Impact of KK Fermions on $Zd_i\bar d_j$ Couplings}
\unboldmath
\label{sec:Z_d_d}
In Sections~\ref{sec:non-diag_Z_d} and \ref{sec:non-diag_Z_u} we gave the flavour off-diagonal entries in the coupling of $Z$ to up- and down-quarks stemming from the mixing of neutral gauge bosons.
We will next evaluate (\ref{eq:A_L_Z}) and (\ref{eq:A_R_Z}) in the model considered in~\cite{Albrecht:2009xr},
that comprise the effects of mixing of SM quarks with KK-quarks. Both these effects that lead to flavour off-diagonal entries in the Z coupling matrices are of order $\mathcal O(v^2/M_\text{KK}^2)$ and can hence be analysed separately.
To evaluate (\ref{eq:A_L_Z}) and (\ref{eq:A_R_Z})
we recall the specific form of the fields (\ref{eq:2.3}) and (\ref{eq:2.4}) in the notation of~\cite{Albrecht:2009xr},
\bea
\Psi_L^T(-1/3)&=&\left(q_L^{d_i(0)},\,q_L^{d_i},\,D_L^{\prime i},\,D_L^i\right)\label{eq:Psi_L_RS}\,,\\
\Psi_R^T(-1/3)&=&\left(D_R^{i(0)},\,q_R^{d_i},\,D_R^{\prime i},\,D_R^i\right)\,.\label{eq:Psi_R_RS}
\eea
The non-vanishing block-matrices $\left[A_L^{-1/3}(Z)\right]_{kj}$, $\left[A_R^{-1/3}(Z)\right]_{kj}$ and $\left[A_{L,R}^{-1/3}(Z)\right]_{00}$ are collected in Table~\ref{tab:A_L_R}. All these matrices are proportional to the $3\times3$ unit matrix and we list only the overall factors that are flavour independent and represent the relevant weak charges that can be easily computed using equation (\ref{eq:g}) together with the quantum numbers of Table~\ref{tab:fieldcontent} and are given in the Appendix. The vector-like couplings of heavy fermions should be noted. 
\begin{table}
\centering
\renewcommand{\arraystretch}{1.5}
\begin{tabular}{|c|c|c|c|c|}
\hline
&(0,0)&(1,1)&(2,2)&(3,3)\\\hline
$A_L^{-1/3}$&$g^{4D}_{Z,L}(d)$&$g^{4D}_{Z,L}(d)$&$g^{4D}_Z(D^{\prime})$&$g^{4D}_{Z,R}(d)$\\\hline
$A_R^{-1/3}$&$g^{4D}_{Z,R}(d)$&$g^{4D}_{Z,L}(d)$&$g^{4D}_Z(D^{\prime})$&$g^{4D}_{Z,R}(d)$\\\hline
\end{tabular}
\renewcommand{\arraystretch}{1.0}
\caption{Weak charges in the coupling matrices of down-quarks to the $Z$ gauge boson.\label{tab:A_L_R}}
\end{table}
Using (\ref{eq:A_L_Z}) we first find 
\bea
A_L^{-1/3}(Z)&=&g^{4D}_{Z,L}(d)\mathbbm{1}\nonumber\\
&+&\left(g^{4D}_Z(D^{\prime})-g^{4D}_{Z,L}(d)\right)M_{02}\frac{1}{M_2^2}M_{02}^\dagger\nonumber\\
&+&\left(g^{4D}_{Z,R}(d)-g^{4D}_{Z,L}(d)\right)M_{03}\frac{1}{M_3^2}M_{03}^\dagger\,,
\eea
where for the mass matrix elements $M_{ij}=M_{ij}(-1/3)$ are given in (4.16) of \cite{Albrecht:2009xr}.
Evidently, the terms involving $M_1$ cancelled each other as a consequence of
$\left[ A_L^{-1/3}\right]_{00}=\left[A_L^{-1/3}\right]_{11}$. With (\ref{eq:App1}), (\ref{eq:App2}) and (\ref{eq:g_Z}) we finally find
\be
A_L^{-1/3}(Z)=g^{4D}_{Z,L}(d)\mathbbm{1}+\frac{1}{2}\frac{g^{4D}}{\cos\psi}\left(M_{03}\frac{1}{M_3^2}M_{03}^\dagger-M_{02}\frac{1}{M_2^2}M_{02}^\dagger\right)\,.
\label{eq:A_L_Z_down_RS}
\ee
In the limit {of $P_{LR}$ being an exact symmetry $P_{LR}(D)=D^\prime$ holds and as a consequence we have} $\left|M_{03}\right|=\left|M_{02}\right|$, $M_3=M_2$ which guarantees that the $\mathcal{O}(v^2/M_\text{KK}^2)$ correction to the coupling $A_L^{2/3}(Z)$ vanishes, expressing the protection of $Zd_L^i\bar d_L^j$ in the presence of mixing with KK fermions.

On the other hand using (\ref{eq:A_R_Z}) we first find
\bea
A_R^{-1/3}(Z)&=&g^{4D}_{Z,R}(d)\mathbbm{1}\nonumber\\
&+&\left(g^{4D}_{Z,L}(d)-g^{4D}_{Z,R}(d)\right)M_{10}^\dagger\frac{1}{M_1^2}M_{10}\nonumber\\
&+&\left(g^{4D}_Z(D^{\prime})-g^{4D}_{Z,R}(d)\right)M_{20}^\dagger\frac{1}{M_2^2}M_{20}\,.
\eea
This time the terms involving $M_3$ cancel each other as a consequence of
$\left[A_R^{-1/3}\right]_{00}=\left[A_R^{-1/3}\right]_{33}$. Using (\ref{eq:App1}), (\ref{eq:App2}) and (\ref{eq:g_Z}) we then find
\be
A_R^{-1/3}(Z)=g^{4D}_{Z,R}(d)\mathbbm{1}-\frac{g^{4D}}{\cos\psi}\left(\frac{1}{2}M_{10}^\dagger\frac{1}{M_1^2}M_{10}+M_{20}^\dagger\frac{1}{M_2^2}M_{20}\right)\,.
\label{eq:A_R_Z_down_RS}
\ee
Now the terms $\mathcal{O}(v^2/M_\text{KK}^2)$ do not cancel each other and the mixing of SM quarks with KK fermions has an impact on right-handed down-quark couplings to $Z$.

\boldmath
\subsection{Impact of KK Fermions on $Zu_i\bar u_j$ Couplings}\label{Sec:Z_u_u}
\unboldmath
In this section, analogous to Section~\ref{sec:Z_d_d}, we evaluate (\ref{eq:A_L_Z}) and (\ref{eq:A_R_Z}) in the model considered in~\cite{Albrecht:2009xr}, this time for the couplings to up-quarks\footnote{The protection of the right-handed up-quark diagonal coupling by the $P_{LR}$ symmetry has been previously considered in \cite{Atre:2008iu}.}. The specific form of the fields (\ref{eq:2.1}) and (\ref{eq:2.2}) in the notation of~\cite{Albrecht:2009xr} is given by
\be\label{eq:Psi_L23}
\Psi_L^T(2/3)=\left(q_L^{u_i(0)},q_L^{u_i},U_L^{\prime i},U_L^{\prime\prime i},\chi_L^{d_i},u_L^i\right)\,,
\ee
\be\label{eq:Psi_R23}
\Psi_R^T(2/3)=\left(u_R^{i(0)},q_R^{u_i},U_R^{\prime i},U_R^{\prime\prime i},\chi_R^{d_i},u_R^i\right)\,.
\ee
The non-vanishing $\left[A_R^{2/3}(Z)\right]_{kj}$, $\left[A_R^{2/3}(Z)\right]_{kj}$ and $\left[A^{2/3}_{L,R}(Z)\right]_{00}$ are collected in Table~\ref{tab:A_L_R_up}. Again, all these matrices are proportional to the $3\times3$ unit matrix and we list only the overall factors that are flavour independent and represent the relevant weak charges. The vector-like couplings of heavy fermions {again can be obtained} using equations (\ref{eq:g}) together with Table~\ref{tab:fieldcontent} and are given in the Appendix.

Using (\ref{eq:A_R_Z}) we first find
\bea
A_R^{2/3}(Z)&=&g^{4D}_{Z,R}(u)\mathbbm{1}\nonumber\\
&+&\left(g^{4D}_{Z,L}(u)-g^{4D}_{Z,R}(u)\right)M_{10}^\dagger \frac{1}{M_1^{2}}M_{10}\nonumber\\
&+&\left(g^{4D}_Z(\chi^d)-g^{4D}_{Z,R}(u)\right)M_{40}^\dagger \frac{1}{M_4^{2}}M_{40}\,,
\eea
where this time the mass matrix elements $M_{ij}=M_{ij}(2/3)$ are given in (4.15) of~\cite{Albrecht:2009xr}.
Note that the terms in the above expression are related by the custodial parity $P_{LR}$, which acts on the quark fields as $P_{LR}(q^u)=\chi^d$, $P_{LR}(u)=u$, and also ensures that $\left|M_{10}\right|=\left|M_{40}\right|$ and $M_1=M_4$, up to small symmetry breaking effects by the BCs on the UV brane.
With {the explicit charge factors given in the Appendix} we finally find
\be
A_R^{2/3}(Z)=g^{4D}_{Z,R}(u)\mathbbm{1}+\frac{1}{2}\frac{g^{4D}}{\cos\psi}\left(M_{10}^\dagger \frac{1}{M_1^{2}}M_{10}-M_{40}^\dagger \frac{1}{M_4^{2}}M_{40}\right)\,.
\label{eq:A_R_Z_up_RS}
\ee
In the limit $\left|M_{10}\right|=\left|M_{40}\right|$, $M_1=M_4$, the $\mathcal O(v^2/M_\text{KK}^2)$ correction to the coupling $A_R^{2/3}(Z)$ vanishes, expressing the protection of $Zu_R^i\bar u_R^j$ in the presence of mixing with KK fermions.

On the other hand, using (\ref{eq:A_L_Z}) we first find
\bea
\label{eq:A_L_Z_up}
A_L^{2/3}(Z)&=&g^{4D}_{Z,L}(u)\mathbbm{1}\nonumber\\
&+&\left(g^{4D}_Z(U^\prime)-g^{4D}_{Z,L}(u)\right)M_{02}\frac{1}{M_2^{2}}M_{02}^\dagger\nonumber\\
&+&\left(g^{4D}_Z(U^{\prime\prime})-g^{4D}_{Z,L}(u)\right)M_{03}\frac{1}{M_3^{2}}M_{03}^\dagger\nonumber\\
&+&\left(g^{4D}_{Z,R}(u)-g^{4D}_{Z,L}(u)\right)M_{05}\frac{1}{M_5^{2}}M_{05}^\dagger\,.
\eea
This time the terms in (\ref{eq:A_L_Z_up}) are not related by the custodial 
parity $P_{LR}$. Using {the explicit charge factors given in the Appendix} we find then
\be
A_L^{2/3}(Z)=g^{4D}_{Z,L}(u)\mathbbm{1}-\frac{1}{2}\frac{g^{4D}}{\cos\psi}\left(M_{02}\frac{1}{M_2^{2}}M_{02}^\dagger+M_{03}\frac{1}{M_3^{2}}M_{03}^\dagger+M_{05}\frac{1}{M_5^{2}}M_{05}^\dagger\right)\,.
\label{eq:A_L_Z_up_RS}
\ee
Now the terms $\mathcal O(v^2/M_\text{KK}^2)$ do not cancel each other and the mixing of SM quarks with KK fermions has an impact on the left-handed up-quark couplings to $Z$.
\begin{table}
\centering
\renewcommand{\arraystretch}{1.5}
\begin{tabular}{|c|c|c|c|c|c|c|}
\hline
&(0,0)&(1,1)&(2,2)&(3,3)&(4,4)&(5,5)\\\hline
$A_L^{2/3}$&$g^{4D}_{Z,L}(u)$&$g^{4D}_{Z,L}(u)$&$g^{4D}_Z(U^\prime)$&$g^{4D}_Z(U^{\prime\prime})$&$g^{4D}_Z(\chi^d)$&$g^{4D}_{Z,R}(u)$\\\hline
$A_R^{2/3}$&$g^{4D}_{Z,R}(u)$&$g^{4D}_{Z,L}(u)$&$g^{4D}_Z(U^\prime)$&$g^{4D}_Z(U^{\prime\prime})$&$g^{4D}_Z(\chi^d)$&$g^{4D}_{Z,R}(u)$\\\hline
\end{tabular}
\renewcommand{\arraystretch}{1.0}
\caption{Weak charges in the coupling matrices of up-quarks to the $Z$ gauge boson.\label{tab:A_L_R_up}}
\end{table}

\subsection{Higgs Couplings}\label{sec:Higgscouplings}
We next evaluate the corrections to the off-diagonal couplings of the Higgs boson with down-quarks in (\ref{eq:Y_non-diag}) using the mass matrix (4.16) of~\cite{Albrecht:2009xr}. 
{Starting with the first two terms in (\ref{eq:Y_non-diag}) corresponding to $Y_1$ in (\ref{eq:Y_1_2}) we find}
\bea
\left[Y_\text{mass}^{-1/3}\right]^{(1)}_\text{non-diag}&=&2\mathcal D_L^\dagger\left(v^2Y_{02}\frac{1}{M_2}Y_{21}^\dagger\frac{1}{M_1}Y_{10}\right)\mathcal D_R\nonumber\\
&+&2\mathcal D_L^\dagger\left(v^2Y_{03}\frac{1}{M_3}Y_{31}^\dagger\frac{1}{M_1}Y_{10}\right)\mathcal D_R\,.
\eea
However in the case of the Higgs being localised on the IR brane $Y_{21}^\dagger=Y_{31}^\dagger=0$ and there is no correction from the mixing with KK fermions {to $Y_1$ at $\mathcal{O}(v^2/M_\text{KK}^2)$}.

The Higgs coupling to the up-quarks can be evaluated in an analogous way, this time using the mass matrix (4.15) of~\cite{Albrecht:2009xr}. 
{The contribution to $Y_1$ reads}
\bea
\left[Y_\text{mass}^{2/3}\right]^{(1)}_\text{non-diag}&=&2\mathcal U_L^\dagger\left(v^2Y_{02}\frac{1}{M_2}Y_{21}^\dagger\frac{1}{M_1}Y_{10}\right)\mathcal U_R
+2\mathcal U_L^\dagger\left(v^2Y_{03}\frac{1}{M_3}Y_{34}^\dagger\frac{1}{M_4}Y_{40}\right)\mathcal U_R\nonumber\\
&+&2\mathcal U_L^\dagger\left(v^2Y_{02}\frac{1}{M_2}Y_{24}^\dagger\frac{1}{M_4}Y_{40}\right)\mathcal U_R
+2\mathcal U_L^\dagger\left(v^2Y_{03}\frac{1}{M_3}Y_{31}^\dagger\frac{1}{M_1}Y_{10}\right)\mathcal U_R\nonumber\\
&+&2\mathcal U_L^\dagger\left(v^2Y_{05}\frac{1}{M_5}Y_{51}^\dagger\frac{1}{M_1}Y_{10}\right)\mathcal U_R
+2\mathcal U_L^\dagger\left(v^2Y_{05}\frac{1}{M_5}Y_{54}^\dagger\frac{1}{M_4}Y_{40}\right)\mathcal U_R\,.\nonumber\\
\eea
As above, in the case of the Higgs being localised on the IR brane some elements of the mass matrix vanish, $Y_{21}^\dagger=Y_{31}^\dagger=Y_{51}=Y_{24}^\dagger=Y_{34}^\dagger=Y_{54}^\dagger=0$, and there is no correction from the mixing with KK fermions {to $Y_1$ at order $\mathcal O(v^2/M_\text{KK}^2)$ in agreement with~\cite{Blanke:2008zb}. 
The remaining non-zero contribution to $Y$ resulting from $Y_2$ is $\mathcal O(vm_i/M_\text{KK}^2)$ again in agreement with~\cite{Blanke:2008zb}, where $m_i$ is the mass of the involved light quark.}\\

Strictly speaking, the above result is only valid in the limit where the couplings 
\begin{equation}
\bar q_R Y_d^\prime D_L \phi+\bar q_R Y_u^\prime u_L
\end{equation}
\noindent are set to zero. These couplings are not required for the generation of SM masses and, in the case of brane Higgs, are a priori unrelated to the SM-like Yukawa matrices, consequently it is possible to set them to zero. \footnote{However, as stated by Azatov et al.~in \cite{Azatov:2009na}, this approach is against the general naturalness argument according to which all dimensionless 5D parameters are expected to be of order one.}

If the Yukawa couplings $Y_u^\prime$ and $Y_d^\prime$ are included into the analysis the situation indeed changes. It was pointed out recently in \cite{Azatov:2009na} that the profiles of the quark fields with Dirichlet BC on the IR brane do not vanish after EWSB but display a discontinuity that is proportional to the brane Higgs vacuum expectation value. After regularisation this induces a very tiny but finite overlap with the Higgs boson even in the case of a brane localised Higgs. Consequently,
the sum over the infinite KK tower of fermion modes can lead to a sizable contribution to Higgs FCNCs, especially in the case of a light Higgs boson.

\newsection{Impact of Gauge Boson Mixing and KK Fermions on Charged Couplings\label{sec:Impact_charged}}
\subsection{Preliminaries}
In this section using the formulae (\ref{eq:G_L_W}) and (\ref{eq:G_R_W}) we will investigate the impact of KK fermions on the charged current processes. As the mixing of $W^\pm$ with the heavy charged gauge bosons $W_H$ and $W^\prime$ has also an impact on these processes we will work out a number of formulae for this case as well, in order to be able to compare the size of these two different effects numerically in Section~\ref{sec:Numerics}. The highlights of this section are the breakdown of the unitarity of the CKM matrix and the generation of the $W^\pm$ couplings to right-handed quarks. It should be emphasised that charged current processes are not protected by the custodial symmetries discussed in the previous section. But as the charged current processes take place in the SM already at tree level such a protection is less important except possibly for the $W^+\bar t_L b_L$ coupling, for which the effect turns out to be largest.

\subsection{Breakdown of Unitarity of the CKM Matrix\label{sec:breakdown_of_unitarity}\label{sec:7_2}}
\subsubsection{Preliminaries}
When discussing the breakdown of unitarity of the CKM matrix one 
should emphasise that when $\mathcal O(v^2/M^2_\text{KK})$ corrections to the $W^+\bar q_iq_j$
vertex are present in a given model they can either be considered as 
contributions to the CKM matrix or treated separately as new effective
flavour and CP-violating charged current interactions. 
In the latter case the CKM matrix remains clearly unitary and the
$\mathcal O(v^2/M^2_\text{KK})$ corrections to the $W^+\bar q_iq_j$ vertex 
that otherwise would modify the CKM matrix give additional explicit 
contributions to the decay amplitudes. Clearly the physical results
for decay amplitudes do not depend on whether the corrections in question 
have been added to the CKM matrix or treated separately.

In the present paper that has as the main goal the study of the impact of
     KK fermions on SM vertices, we found it more convenient to 
     include all $\mathcal O(v^2/M^2_\text{KK})$ corrections to the $W^+\bar q_iq_j$ vertex into the CKM matrix.

 Another related issue is the definition of the gauge coupling constant,
      denoted by $g^{4D}$. Its value is usually determined with the
      help of the muon decay that also receives $\mathcal O(v^2/M^2_\text{KK})$ corrections
      in the model in question modifying the extracted numerical value of
      $g^{4D}$. While being aware of the presence of such effects, their
      analysis would necessarily require the study of electroweak precision
      observables that in most cases have nothing to do with flavour violation.
      Therefore we decided to postpone the study of such effects to a future
      publication.

{With our definition of the CKM matrix, its unitarity is broken in the model
 under consideration in two ways, namely by:}  
\begin{itemize}
 \item The non-universality of gauge interactions of $W^\pm$ with SM fermions through its mixing with the heavy gauge bosons in the process of EWSB.
\item The mixing of the SM fermions with the heavy KK fermions.
\end{itemize}
Both effects are $\mathcal{O}(v^2/M_\text{KK}^2)$ and it is of interest to see how large these two effects are, whether one of them is dominant and whether this dominance is flavour dependent. At $\mathcal{O}(v^2/M_\text{KK}^2)$ one can consider these two effects independently from each other, which we will do in the following.

\boldmath
\subsubsection{Non-Universality in the Couplings of $W^+$\label{sec:nonuniversality}}
\unboldmath
We consider the charged current in (\ref{eq:J_mu_W}) after EWSB with $W^+$ denoting the mass eigenstate found after the diagonalisation of the $3\times 3$ gauge boson mass matrix as done in \cite{Albrecht:2009xr}. From the Table~13 of the latter paper we learn that
\begin{equation}\label{eq:G_L}
 \left[G_L(W^+)\right]_{00}\equiv G_L(W^+)=\frac{g^{4D}}{\sqrt{2}}\left(\mathbbm{1}+\frac{v^2}{M_\text{KK}^2}\Delta_\text{G}\right)\,,
\end{equation}
where 
\begin{equation}
 \Delta_\text{G}=-\frac{\left(g^{4D}\right)^2}{4}\mathcal{I}_1^+\underset{00}{\mathcal{R}_1^i}(++)\,.
\end{equation}
Here $\mathcal{I}_1^+$ is flavour independent, while $\underset{00}{\mathcal{R}_1}(++)$ is a diagonal $3\times 3$ matrix in flavour space,
\begin{equation}
 \underset{00}{\mathcal{R}_1}(++)=\left(\underset{00}{\mathcal{R}_1^1}(++),\underset{00}{\mathcal{R}_1^2}(++),\underset{00}{\mathcal{R}_1^3}(++)\right)\,,
\end{equation}
with {its diagonal elements} $\underset{00}{\mathcal{R}_1^i}(++)$ given in (\ref{eq:intR}).

The dependence of $\underset{00}{\mathcal{R}_1^i}(++)$ on the flavour index ``$i$'' signals the breakdown of flavour universality in gauge interactions of quark flavour eigenstates and as we will soon see, the breakdown of unitarity of the CKM matrix.

The rotation to quark mass eigenstates in the absence of the mixing with KK fermions by means of unitary matrices $\mathcal U_{L,R}$ and $\mathcal D_{L,R}$ transforms $G_L(W^+)$ in (\ref{eq:G_L}) into
\begin{equation}\label{eq:CKM}
 \mathcal U_L^\dagger G_L(W^+)\mathcal D_L=\frac{g^{4D}}{\sqrt{2}}V_\text{CKM}^\text{G}\,,
\end{equation}
where we defined 
\begin{equation}\label{eq:CKMKKgauge}
 V_\text{CKM}^\text{G}=V_\text{CKM}^0+\frac{v^2}{M^2}\mathcal U_L^\dagger\Delta_\text{G}\mathcal D_L\,,
\end{equation}
with $V_\text{CKM}^0=\mathcal U_L^\dagger\mathcal D_L$ being a unitary matrix. As the $\mathcal O(v^2/M_\text{KK}^2)$ correction is non-unitary, $V_\text{CKM}^\text{G}$ is non-unitary as well.

\subsubsection{Mixing with KK Fermions}\label{sec:violationfermions}
Let us denote the result in (\ref{eq:G_L_W}) in analogy to (\ref{eq:G_L}) by
\begin{equation}\label{eq:G_LKK}
 G_L^\text{KK}(W^+)=\frac{g^{4D}}{\sqrt{2}}\left(\mathbbm{1}+\frac{v^2}{M_\text{KK}^2}\Delta_\text{KK}\right)\,,
\end{equation}
where $\Delta_\text{KK}$ is a non-diagonal $3\times 3$ matrix defined through (\ref{eq:G_L_W}) and (\ref{eq:G_LKK}). 
In (\ref{eq:M_diag_down}) and (\ref{eq:M_diag_up}) we diagonalised the effective mass matrix (\ref{eq:mass-matrix}) for $-1/3$ and $+2/3$ quarks by $\mathcal D_{L,R}$ and $\mathcal U_{L,R}$, respectively. Accordingly we now find
\be\label{eq:CKMKKfermions}
V_\text{CKM}^\text{KK}=V_\text{CKM}^0+\frac{v^2}{M^2}\mathcal U_L^\dagger\Delta_\text{KK}\mathcal D_L\,.
\ee
The $\mathcal{O}(v^2/M_\text{KK}^2)$ correction {in (\ref{eq:CKMKKfermions}) breaks the unitarity of the CKM matrix.}

\subsubsection{Testing the Breakdown of Unitarity\label{sec:Testing_Breakdown}}
Let us denote the matrices $\Delta_\text{G}$ and $\Delta_\text{KK}$ by a common symbol $\Delta_r$. Then one can easily find for each contribution corrections to the standard relation for the unitarity of the CKM matrix,
\begin{eqnarray}
K^u_r\equiv
V_\text{CKM}^rV_\text{CKM}^{^r\dagger}&=&\mathbbm{1}+\frac{v^2}{M_\text{KK}^2}\mathcal
U_L^\dagger\left(\Delta_r+\Delta_r^\dagger\right)\mathcal U_L\,, \label{eq:UT1}\\
K^d_r \equiv V_\text{CKM}^{r\dagger} V_\text{CKM}^r&=&\mathbbm{1}+\frac{v^2}{M_\text{KK}^2}\mathcal D_L^\dagger\left(\Delta_r+\Delta_r^\dagger\right)\mathcal D_L\,.\label{eq:UT2}
\end{eqnarray}
One can now test how {the twelve usual unitarity relations are violated. We will return to this issue and numerically investigate the breakdown of unitarity in Section~\ref{sec:violationnumerics}}.

\boldmath
\subsection{The Mixing Matrix for Right-Handed Quarks}
\unboldmath
In the SM the $W^\pm$ gauge boson couples only to left-handed quarks. This
property is not modified through the mixing of $W^\pm$ with new heavy charged 
gauge bosons. On the other hand the mixing of SM quarks with KK fermions 
generates non-zero couplings of $W^\pm$ to right-handed quarks. The relevant
formulae are given in (\ref{eq:G_R_W}) and (\ref{eq:G_W_mass_eigenstates}).
In analogy to {the CKM matrix defined in} (\ref{eq:CKM}) we can now define the matrix $V_\text{R}$ through
\begin{equation}\label{eq:VR}
 \mathcal U_R^\dagger G_R(W^+)\mathcal D_R=\frac{g^{4D}}{\sqrt{2}}V_\text{R}\,.
\end{equation}
The matrix $V_\text{R}$ describes the pattern of flavour violation in the 
charged right-handed currents. It is of interest to investigate whether 
there is a hierarchy in the elements of this matrix and in such case how
does it compare to the hierarchy in the elements of the CKM matrix. We will return to this question in Section~\ref{sec:Numerics}.

\newsection{Numerical Analysis\label{sec:Numerics}}
\subsection{Preliminaries}
The expressions that we have found for the effective couplings of SM quarks to gauge bosons and the Higgs boson (\ref{eq:A_Z_down_mass_eigenstates})-(\ref{eq:Y_H_up_mass_eigenstates}) together with (\ref{eq:A_L_Z})-(\ref{eq:Y_H}) and (\ref{eq:Y_non-diag})
are valid up to $\mathcal{O}(v^2/M_\text{KK}^2)$. To get a feeling for the actual accuracy of these formulae, in this section we will compare our results to the full numerical computation in the RS model analysed in \cite{Albrecht:2009xr,Blanke:2008zb,Blanke:2008yr}.
Having established the accuracy of the coupling matrices for the off-diagonal couplings of the gauge bosons and the Higgs derived in the effective theory, we will investigate
\begin{itemize}
 \item the size of the corrections to the $Z$ couplings
 \item the violation of the CKM unitarity as described in Section~\ref{sec:Testing_Breakdown}
 \item the impact of $\mathcal O(v^2/M_\text{KK}^2)$ corrections on the actual values of the CKM matrix entries
 \item the structure of the right-handed mixing matrix $V_\text{R}$ generated through the mixing of SM fermions with the KK fermions.
\end{itemize}

While analysing the first three points on our list above we will compare the impact of mixing between SM fermions and the lightest KK fermions with the impact of gauge boson mixing and identify the dominant contribution.

\subsection{Accuracy of Effective Theory Expressions\label{sec:Accuracy}}
\subsubsection{Numerical Strategy\label{sec:Numerical_Strategy}}
The numerical calculation of the couplings of the $Z$, $W^\pm$ and Higgs to SM quarks is straightforward. Starting from the (diagonal) coupling matrices $\mathcal{A}_{L,R}^{Q}(Z)$, $\mathcal{G_{L,R}(W^+)}$ and (off-diagonal) $\mathcal{Y}(Q)$ ($Q=-1/3,\,2/3$) {given in equations (\ref{eq:Yukawa})-(\ref{eq:J_mu_W})} the mixing with KK-fermions is incorporated by rotating these coupling matrices to the mass eigenbasis. This is achieved by four unitary matrices, $\mathcal{U}_{L,R}^{18}$ and $\mathcal{D}_{L,R}^{12}$, that are $18\times18$ and $12\times12$ matrices, respectively. They are defined through
\be
\mathcal{M}_\text{diag}(2/3)=\left(\mathcal{U}_L^{18}\right)^\dagger\mathcal{M}(2/3)\ \mathcal{U}_R^{18}\,,\qquad 
\mathcal{M}_\text{diag}(-1/3)=\left(\mathcal{D}_L^{12}\right)^\dagger\mathcal{M}(-1/3)\ \mathcal{D}_R^{12}\,,
\ee
where $\mathcal{M}(2/3)$ and $\mathcal{M}(-1/3)$, as defined in (\ref{eq:mass-terms}), are given in equations (4.15), (4.16) of \cite{Albrecht:2009xr} and have dimensions $18\times18$ and $12\times12$, respectively. Having determined these unitary matrices, the effective coupling matrices in the mass eigenstate basis can be calculated analogous to (\ref{eq:A_Z_down_mass_eigenstates})
-(\ref{eq:Y_H_up_mass_eigenstates}). The {coupling matrices for SM quarks} then are given by the upper-left $3\times3$ blocks of the resulting coupling matrices in the mass eigenstate basis. The origin of off-diagonal entries in these $3\times3$ sub-matrices in the case of {gauge couplings} is the non-universality of the full coupling matrices due to different weak charges of the involved fermions. In the case of the {Higgs couplings} the origin lies in the fact that the Higgs coupling matrices, $\mathcal{Y}(Q)$, $Q=2/3,-1/3$, have in contrast to the corresponding mass matrices $\mathcal{M}(Q)$ no diagonal terms of order $\mathcal{O}(M_\text{KK})$, and hence the matrices $\mathcal{Y}(Q)$ and $\mathcal{M}(Q)$ cannot be diagonalised simultaneously.

The numerical diagonalisation of matrices as large as $\mathcal{M}(2/3)$ and $\mathcal{M}(-1/3)$ is numerically involved and time consuming, and beyond that the large hierarchies present in these matrices - diagonal entries have absolute sizes ranging from $\mathcal O(m_u)$ to $\mathcal O(M_\text{KK})$ - limit the accuracy of the diagonalisation. Although a numerical calculation of the KK-quark impact on the coupling matrices is in principle feasible, it would on that account be preferable to have confidence in the accuracy of expressions (\ref{eq:A_Z_down_mass_eigenstates})-(\ref{eq:Y_H_up_mass_eigenstates}) derived in the effective theory and be able to use them instead of the full expressions.

\subsubsection{Results\label{sec:Results}}
\begin{figure}
\center{\includegraphics[width=.4\textwidth]{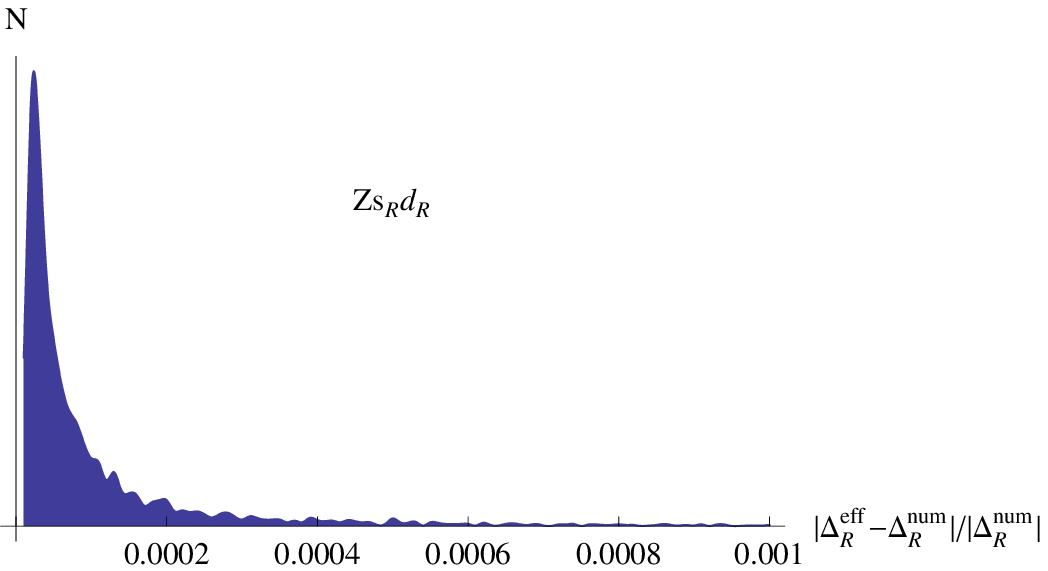}\hspace{.5cm}\includegraphics[width=.4\textwidth]{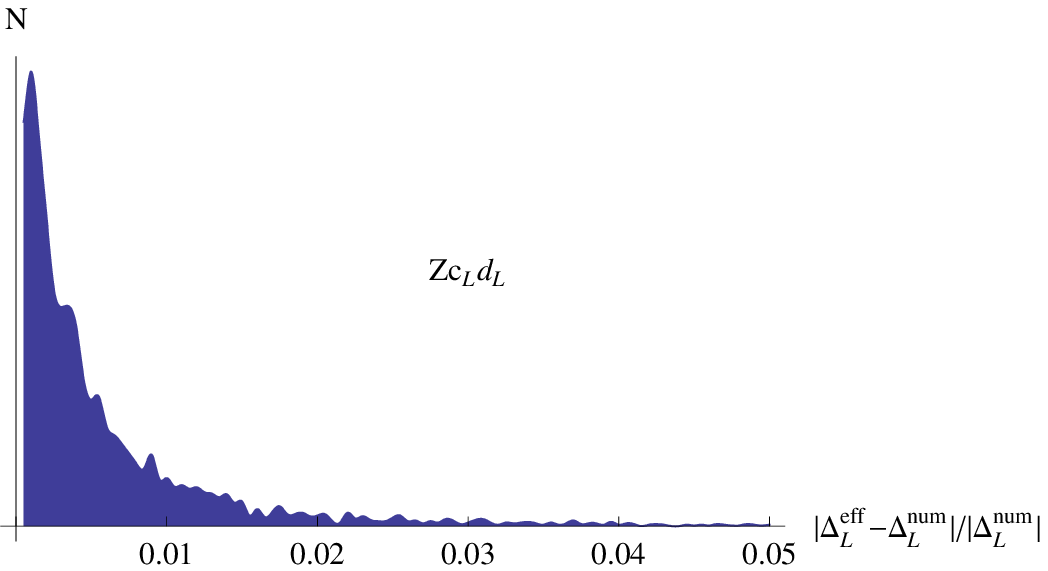}}
\caption{Comparison of the effective theory result to the result of the full calculation for the $Zs_R \bar d_R$ coupling (left panel) and $Zc_L \bar u_L$ coupling (right panel).\label{fig:accuracy_Z}}
\end{figure}

We have compared the KK fermion  corrections to the couplings of the SM quarks to the $Z$ and the $W^+$ gauge bosons obtained
 by means of the effective theory approach with those obtained by means
 of exact diagonalisation as outlined in Section~\ref{sec:Numerical_Strategy}. Our results
 can be summarised briefly as follows:
\begin{itemize}
\item
The corrections to the couplings $Z d_R^i \bar d_R^j$ and $Zu_L^i \bar u_L^j$, 
that are not protected by the custodial symmetry are
very well described by the effective theory and the agreement with the
full diagonalisation is better than few $\%$ for all points where these
corrections are non-negligible. The corrections to the couplings $W^+u_L^i \bar d_L^j$ and
$W^+u_R^i \bar d_R^j$ that are also not protected by the custodial symmetry are still described adequately (agreement better than $20\%$ for the majority of the points), although not to the excellent accuracy of the former couplings.
As an example we show in Fig.~\ref{fig:accuracy_Z} the deviation between effective theory result and full calculation for the couplings $Zs_R \bar d_R$ and $Zc_L \bar u_L$.
\item
  The corrections to the couplings $Z d_L^i \bar d_L^j$ and $Zu_R^i \bar u_R^j$, 
 that are protected by custodial symmetry are found to be negligibly small.
 For such small couplings it is not surprising that effective theory
 expressions and the full diagonalisation do not fully agree with each 
 other. We attribute these differences to the accuracy of the full 
 diagonalisation of large mass matrices and believe that the 
 {formulae derived in the effective theory} give a better description of these suppressed corrections.
 As the latter are phenomenologically irrelevant the differences encountered
 here are not important.
\item {The corrections to the couplings $Hd_L^i\bar d_R^j$ as well as $Hu_L^3\bar u_R^{1,2}$, $Hu_L^{1,2}\bar u_R^3$ are described adequately by the expressions derived in the effective theory. For the coupling $Hc_L\bar u_R$ which is typically significantly smaller than the other off-diagonal Higgs couplings the effective theory approach reproduces the full calculation up to $\mathcal O(1)$ factors.} 
\end{itemize}

\boldmath
\subsection{The Size of Corrections to the $Z$-Couplings\label{sec:Size_of_Corrections_to_Z}}
\unboldmath
We will next compare the effect of KK fermion mixing on the 
SM couplings to the one arising from the {mixing of gauge bosons}.

To simplify the discussion we denote the contribution from gauge boson mixing as $\Delta^{ij}_\text{G}$ and the contributions from KK-fermion mixing as $\Delta^{ij}_\text{KK}$ as already done in Section~\ref{sec:breakdown_of_unitarity}. For example for the $Zd^i_R\bar d^j_R$ coupling we define
\begin{eqnarray}
A_{R,\text{G}}^{-1/3}(Z)=g^{4D}_{Z,R}(d)\left(\mathbbm{1}+\frac{v^2}{M_\text{KK}^2}\Delta_\text{G}(Z)\right)\,,\\
A_{R,\text{KK}}^{-1/3}(Z)=g^{4D}_{Z,R}(d)\left(\mathbbm{1}+\frac{v^2}{M_\text{KK}^2}\Delta_\text{KK}(Z)\right)\,,
\end{eqnarray}
and analogously for the $Zd^i_L\bar d^j_L$, $Zu^i_{L,R}\bar u^j_{L,R}$ couplings.

\begin{figure}[htbp]
\center{\includegraphics[width=.45\textwidth]{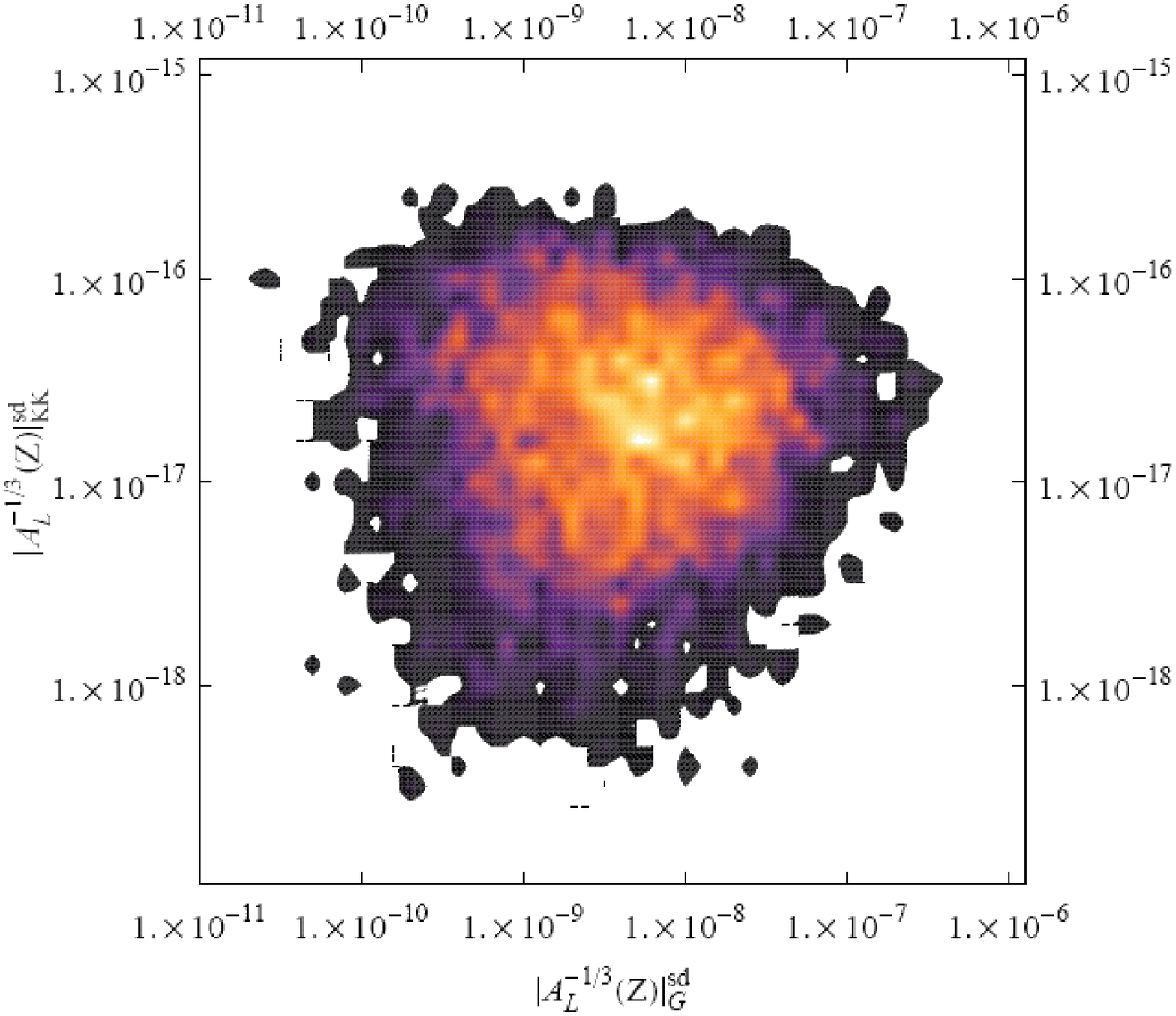}\hspace{.5cm}\includegraphics[width=.45\textwidth]{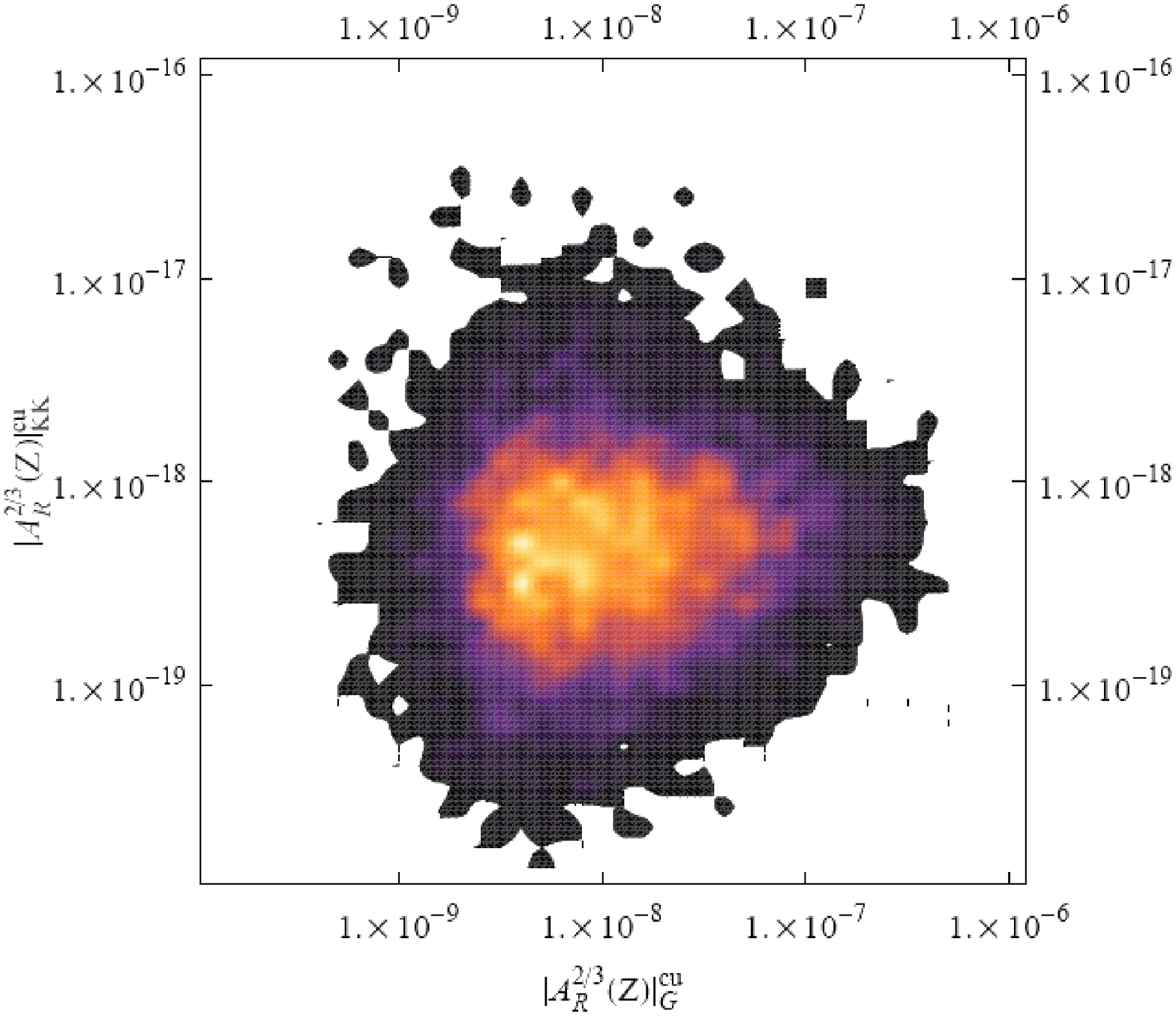}}
\caption{Comparison of contributions from KK-fermion mixing and gauge boson mixing to the custodially protected $Zs_L \bar d_L$ coupling (left panel) an to the custodially protected $Zc_R \bar u_R$ coupling (right panel). {These results have been obtained by using the effective theory expressions.}\label{fig:KK-contrib_Z_P}}
\end{figure}

For the $Zd^i_L \bar d^j_L$ and $Zu^i_R \bar u^j_R$ couplings that are protected by the custodial symmetry the relative impact of KK fermion mixing turns out to be very small. This is due to the fact that the effects of $SU(2)_R\times P_{LR}$ breaking by BCs on the UV brane are much smaller for fermionic KK modes than they are for the gauge boson KK modes. As an example, in Fig.~\ref{fig:KK-contrib_Z_P} we compare the contributions from KK fermion mixing and gauge boson mixing that enter the $Zs_L \bar d_L$ and $Zc_R \bar u_R$ couplings.
In the case of couplings that are not protected by the custodial symmetry, the corrections from KK fermion mixing are 
still subdominant but can in principle be of the same order of magnitude as the contribution from gauge boson mixing. To get a feeling for in which elements of the $Zd_R^i \bar d_R^j$, $Zu_L^i \bar u_L^j$ 
couplings these corrections can potentially become important it is instructive to investigate the patterns of hierarchy in the KK fermion and gauge boson mixing contributions separately and eventually compare them to each other.

We find that the hierarchies in the gauge boson mixing contributions $\Delta_\text{G}(Z)$ that enter the above
couplings of the $Z$ boson are constrained by the presence of the RS-GIM mechanism (cf.~Section~7 in \cite{Blanke:2008yr})\footnote{This basically implies that the hierarchy among flavour off-diagonal couplings $\Delta^{ij}_\text{G}(Z)$ is parallel to the hierarchy in the corresponding mass splittings $|m_i-m_j|$. Details depend on which quark multiplets, e.g. $Q$ or $d$, are involved in the coupling.}.
This should be compared to the flavour hierarchies in the corrections $\Delta_\text{KK}$ 
stemming from KK fermion mixing. From (\ref{eq:Psi_L_RS})-(\ref{eq:Psi_R_RS}), (\ref{eq:Psi_L23})-(\ref{eq:Psi_R23}) and from the mass matrices of up- and down-quarks in~\cite{Albrecht:2009xr} we find that here the patterns are dictated by the hierarchies in the fermion zero mode shape functions on the IR brane, $f^Q$, $f^u$, $f^d$, that are vectors in flavour space and are defined through
\begin{equation}\label{eq:Yud}
Y^{u,d}_{ij} = \lambda^{u,d}_{ij}\,\frac{e^{kL}}{kL} f^{(0)}_{L}(y=L,c_Q^i)  f^{(0)}_{R}(y=L,c_{u,d}^j) \equiv \lambda^{u,d}_{ij}\,\frac{e^{kL}}{kL} f^Q_i  f^{u,d}_j \,.
\end{equation}

In particular, the contributions from KK fermion mixing to gauge couplings should typically be proportional to dyadic products of these quantities, given by e.g.~$\left(f^u\circ f^d\right)_{ij}\equiv f^u_{i}f^d_{j}$. We summarise the expected hierarchies between flavour transitions
for couplings that are not protected by the custodial symmetry
in Table~\ref{tab:hierarchies}.
\begin{table}
\begin{center}
\begin{tabular}{|l|c|c|c|c|}
\hline
&$Zd_Rd_R$&$Zu_Lu_L$&$Wu_Ld_L$&$Wu_Rd_R$\\\hline
$\Delta_\text{G}$&RS-GIM (d)&RS-GIM (Q)&CKM&-\\\hline
$\Delta_\text{KK}$&$f_d\circ f_d$&$f_Q\circ f_Q$&$f_Q\circ f_Q$&$f_u\circ f_d$\\\hline
$\Delta_\text{KK}$ most likely relevant&$sd$&$tc$,$tu$,$cu$&$tb,ts,cb$&all\\\hline
\end{tabular}
\end{center}
\caption{\label{tab:hierarchies}Hierarchies in the gauge boson mixing and KK-fermion mixing contributions to the gauge couplings that are not protected by the custodial symmetry. In the last line we give the elements of the coupling matrices which is on average affected most by KK-fermion mixing.}
\end{table}

\begin{figure}[htbp]
\center{\includegraphics[width=.45\textwidth]{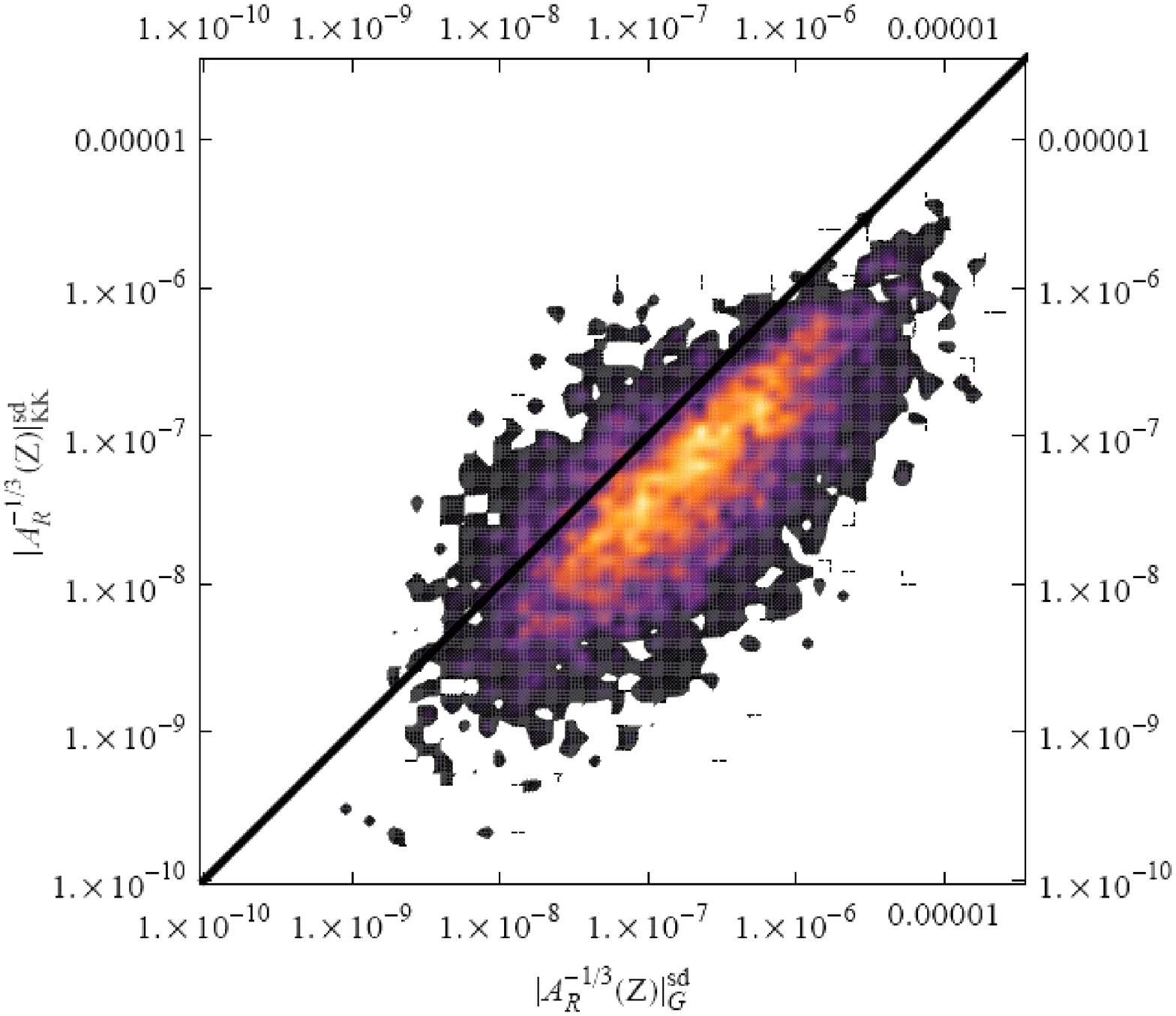}\hspace{.5cm}\includegraphics[width=.45\textwidth]{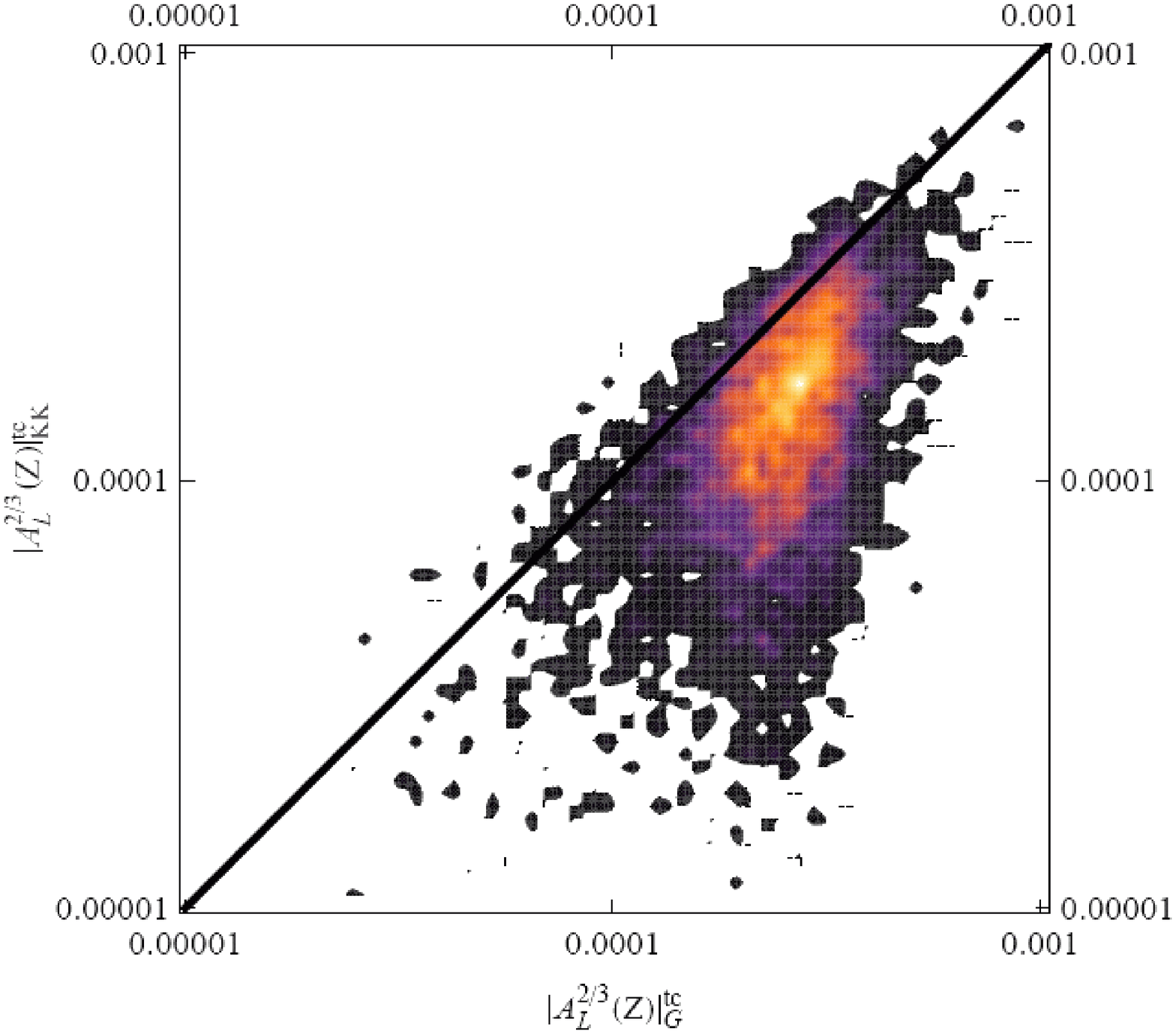}}
\caption{Comparison of contributions from KK-fermion mixing and gauge boson mixing to the unprotected $Zs_R \bar d_R$ coupling (left panel) and to the unprotected $Zt_L \bar c_L$ coupling (right panel).\label{fig:KK-contrib_Z}}
\end{figure}

As indicated above these hierarchies allow to predict which are the gauge couplings where the effects of KK fermion mixing can potentially become important compared to the gauge boson mixing contributions, that is for which flavour transition $j\to i$ the ratio $\Delta_\text{KK}^{ij}/\Delta_\text{G}^{ij}$ between KK fermion mixing and gauge boson mixing effects is on average maximal. 
We list the entries that {receive the largest relative contributions from} KK fermion mixing in the fourth 
{row} of Table~\ref{tab:hierarchies}. In the case of $A_L^{2/3}(Z)$, the flavour hierarchies in gauge boson mixing and KK fermion mixing contributions are roughly equal, such that the relative importance of KK fermion mixing is roughly equal for all flavour transitions of this coupling. 
We compare the gauge boson mixing and KK-fermion mixing contributions for the flavour transitions that are expected to be affected most by the latter contribution in  Fig.~\ref{fig:KK-contrib_Z}.
Also here we find that the contributions from gauge boson mixing typically are the dominant ones.

In summary we find that for all $Z$ couplings the KK-fermion mixing contribution is significantly smaller than the gauge boson mixing contribution for a majority of points in parameter space, and in particular for those points that produce the largest effects in the respective coupling.

\subsection{Violation of CKM Unitarity}\label{sec:violationnumerics}
{We continue our discussion with an analysis of the breakdown of unitarity of the CKM matrix.}
In the class of models we considered above {that contain} $N$ additional charge $2/3$ and $M$ additional charge $-1/3$ vector-like quarks, the CKM matrix generalises to a $3(N+1)\times3(M+1)$ matrix. It is given by
\be
\label{eq:generalized_CKM}
V_\text{CKM}^{N,M}=\left(\mathcal U_L^{3(N+1)}\right)^\dagger\mathcal G_L(W^+)\mathcal D_L^{3(M+1)}\,,
\ee
where $\mathcal U_{L,R}^{3(N+1)}$ and $\mathcal D_{L,R}^{3(M+1)}$ are the
unitary matrices that diagonalise the mass matrices $\mathcal M(2/3)$ and
$\mathcal M(-1/3)$, respectively. The CKM matrix observable at low energies, that is the flavour mixing matrix that is measured in processes with only light quarks as external states, corresponds to the upper-left $3\times3$ block of the generalised CKM matrix and will in general show deviations from unitarity, as indicated in Section~\ref{sec:breakdown_of_unitarity}.
This breakdown {of unitarity} 
can be characterised
 by the departure of the matrices $K^u_r$ and $K^d_r$  in (\ref{eq:UT1}) and
 (\ref{eq:UT2})
    from unit matrices.
 Beginning with the case $r=\text{KK}$ we find that
 \begin{itemize}
 \item
 The corrections to the {\it diagonal} elements of $K^u$ and $K^d$ matrices that
 equal unity in the SM amount to at most $1\%$ for the first two entries. 
 {As seen in Table~\ref{tab:UT_corrections}} for the third generation the corrections can be as large as $2\%$.
 \item
 The corrections to the {\it non-diagonal} elements of $K^u$ and $K^d$ 
 matrices, that vanish in the SM, express the violation of the unitarity
 triangle relations. In Table~\ref{tab:UT_corrections} we list the typical and maximal corrections to the {six} 
 unitarity triangle relations. 
 We give only
 the absolute values of these corrections and compare them in each case 
 to the absolute  value of the terms on the l.h.s.~of the triangle
 relation in question that we evaluate using the actual values of 
 the elements of the CKM matrix. We observe that the relative corrections
 are larger in cases where the quarks of the third generation are involved
 but on the whole the corrections are {below $2\%$ in each case}.
\end{itemize}

For comparison we also analyse the violation of unitarity stemming from gauge
boson mixing, that is for the case $r=G$. Also these results are shown in
  Table~\ref{tab:UT_corrections}. As in the previous case we find the
corrections to the diagonal elements of $K^u_G$ and $K^d_G$ to be at most 1\%
for the first two generations, and at most 2\% for the third generation. For the off-diagonal elements of $K^u$ and $K^d$ we find that corrections
  can be as large as 5\% compared to the largest term on the l.h.s.~of the
  corresponding unitarity relation. Comparing our results to the $r=\text{KK}$
  case above, we find that typically the effects from gauge boson mixing are
  larger than those stemming from KK fermion mixing {except for in the third column and third row unitarity relations, for which the two corrections roughly have the same size.}
\begin{table}[htbp]
\begin{tabular}{|r|c|c|c|c|c|}
\hline
&&$\left\langle K_\text{G}\!-\!\mathbbm{1}\right\rangle$&$|K_\text{G}\!-\!\mathbbm{1}|_\text{max}$&$\left\langle K_\text{KK}\!-\!\mathbbm{1}\right\rangle$&$|K_\text{KK}\!-\!\mathbbm{1}|_\text{max}$\\\hline
$|\vud|^2+|\vcd|^2+|\vtd|^2=$&$K^d_{11}$&$3.5\!\cdot\!10^{-3}$&$3.5\!\cdot\!10^{-3}$&$6.8\!\cdot\!10^{-7}$&$1.9\!\cdot\!10^{-5}$\\
{\color{gray}\small$0.95\quad\,\,\,5\!\cdot\!10^{-2}\quad8\!\cdot\!10^{-5}\quad\!$}&&&&&\\
$|\vus|^2+|\vcs|^2+|\vts|^2=$&$K^d_{22}$&$3.3\!\cdot\!10^{-3}$&$3.5\!\cdot\!10^{-3}$&$2.4\!\cdot\!10^{-5}$&$5.1\!\cdot\!10^{-4}$\\
{\color{gray}\small$5\!\cdot\!10^{-2}\quad0.95\,\,\,\quad2\!\cdot\!10^{-3}\quad\!$}&&&&&\\
$|\vub|^2+|\vcb|^2+|\vtb|^2=$&$K^d_{33}$&$1.4\!\cdot\!10^{-2}$&$1.9\!\cdot\!10^{-2}$&$8.4\!\cdot\!10^{-3}$&$2.1\!\cdot\!10^{-2}$\\
{\color{gray}\small$1\!\cdot\!10^{-5}\quad2\!\cdot\!10^{-3}\quad\,\,\,\,1\qquad\,\,$}&&&&&\\\hline
$|\vud|^2+|\vus|^2+|\vub|^2=$&$K^u_{11}$&$3.5\!\cdot\!10^{-3}$&$3.5\!\cdot\!10^{-3}$&$1.8\!\cdot\!10^{-6}$&$3.3\!\cdot\!10^{-5}$\\
{\color{gray}\small$0.95\,\,\,\,\quad5\!\cdot\!10^{-2}\quad1\!\cdot\!10^{-5}\quad\!$}&&&&&\\
$|\vcd|^2+|\vcs|^2+|\vcb|^2=$&$K^u_{22}$&$3.3\!\cdot\!10^{-3}$&$3.5\!\cdot\!10^{-3}$&$3.9\!\cdot\!10^{-5}$&$4.8\!\cdot\!10^{-4}$\\
{\color{gray}\small$0.95\,\,\,\quad5\!\cdot\!10^{-2}\quad2\!\cdot\!10^{-3}\quad\!$}&&&&&\\
$|\vtd|^2+|\vts|^2+|\vtb|^2=$&$K^u_{33}$&$1.4\!\cdot\!10^{-2}$&$1.9\!\cdot\!10^{-2}$&$8.4\!\cdot\!10^{-3}$&$2.1\!\cdot\!10^{-2}$\\
{\color{gray}\small$8\!\cdot\!10^{-5}\quad\!2\!\cdot\!10^{-3}\,\,\,\,\quad1\qquad\,\,$}&&&&&\\\hline
$\vud\vus^\ast+\vcd\vcs^\ast+\vtd\vts^\ast=$&$K^d_{12}$&$1.4\!\cdot\!10^{-6}$&$5.4\!\cdot\!10^{-5}$&$9.1\!\cdot\!10^{-7}$&$2.5\!\cdot\!10^{-5}$\\
{\color{gray}\small$0.22\qquad\,\,0.22\qquad\!\!4\!\cdot\!10^{-4}\quad$}&&&&&\\
$\vud\vub^\ast+\vcd\vcb^\ast+\vtd\vtb^\ast=$&$K^d_{13}$&$3.7\!\cdot\!10^{-5}$&$3.0\!\cdot\!10^{-4}$&$2.0\!\cdot\!10^{-5}$&$1.8\!\cdot\!10^{-4}$\\
{\color{gray}\small$4\!\cdot\!10^{-3}\quad9\!\cdot\!10^{-3}\quad9\!\cdot\!10^{-3}\quad$}&&&&&\\
$\vus\vub^\ast+\vcs\vcb^\ast+\vts\vtb^\ast=$&$K^d_{23}$&$1.6\!\cdot\!10^{-4}$&$1.6\!\cdot\!10^{-3}$&$9.4\!\cdot\!10^{-5}$&$8.7\!\cdot\!10^{-4}$\\
{\color{gray}\small$9\!\cdot\!10^{-4}\quad4\!\cdot\!10^{-2}\quad4\!\cdot\!10^{-2}\quad$}&&&&&\\\hline
$\vud\vcd^\ast+\vus\vcs^\ast+\vub\vcb^\ast=$&$K^u_{12}$&$1.1\!\cdot\!10^{-5}$&$2.7\!\cdot\!10^{-4}$&$4.5\!\cdot\!10^{-6}$&$1.1\!\cdot\!10^{-4}$\\
{\color{gray}\small$0.22\qquad\,\,0.22\qquad\!\!2\!\cdot\!10^{-4}\quad$}&&&&&\\
$\vud\vtd^\ast+\vus\vts^\ast+\vub\vtb^\ast=$&$K^u_{13}$&$7.2\!\cdot\!10^{-5}$&$4.2\!\cdot\!10^{-4}$&$3.2\!\cdot\!10^{-5}$&$2.2\!\cdot\!10^{-4}$\\
{\color{gray}\small$9\!\cdot\!10^{-3}\quad9\!\cdot\!10^{-3}\quad4\!\cdot\!10^{-3}\quad$}&&&&&\\
$\vcd\vtd^\ast+\vcs\vts^\ast+\vcb\vtb^\ast=$&$K^u_{23}$&$5.9\!\cdot\!10^{-4}$&$1.7\!\cdot\!10^{-3}$&$3.0\!\cdot\!10^{-4}$&$1.1\!\cdot\!10^{-3}$\\
{\color{gray}\small$2\!\cdot\!10^{-3}\quad4\!\cdot\!10^{-2}\quad4\!\cdot\!10^{-2}\quad$}&&&&&\\\hline
\end{tabular}
\caption{CKM unitarity relations and the amount by which they are broken in the RS model. For comparison  in the first column we also give numerical values for the absolute values of the three terms on the l.h.s.~of the relations separately. \label{tab:UT_corrections}}
\end{table}

As seen in Table~\ref{tab:UT_corrections} the effects of CKM unitarity relations coming
 from both the gauge mixing and KK-fermion mixing is very small. 
 Still with improved data one could in principle put some bounds 
 on the parameters of the RS model in question by studying such
 relations. In doing this, as already cautioned in Section~\ref{sec:7_2},
 one would have to carefully study other observables with respect
 to the definition of the gauge coupling at order $v^2/M^2_{KK}$
 and preferably in conjunction with electroweak precision tests.

\subsection{Corrections to the CKM Matrix}
\begin{figure}[htbp]
\center{\includegraphics[width=.45\textwidth]{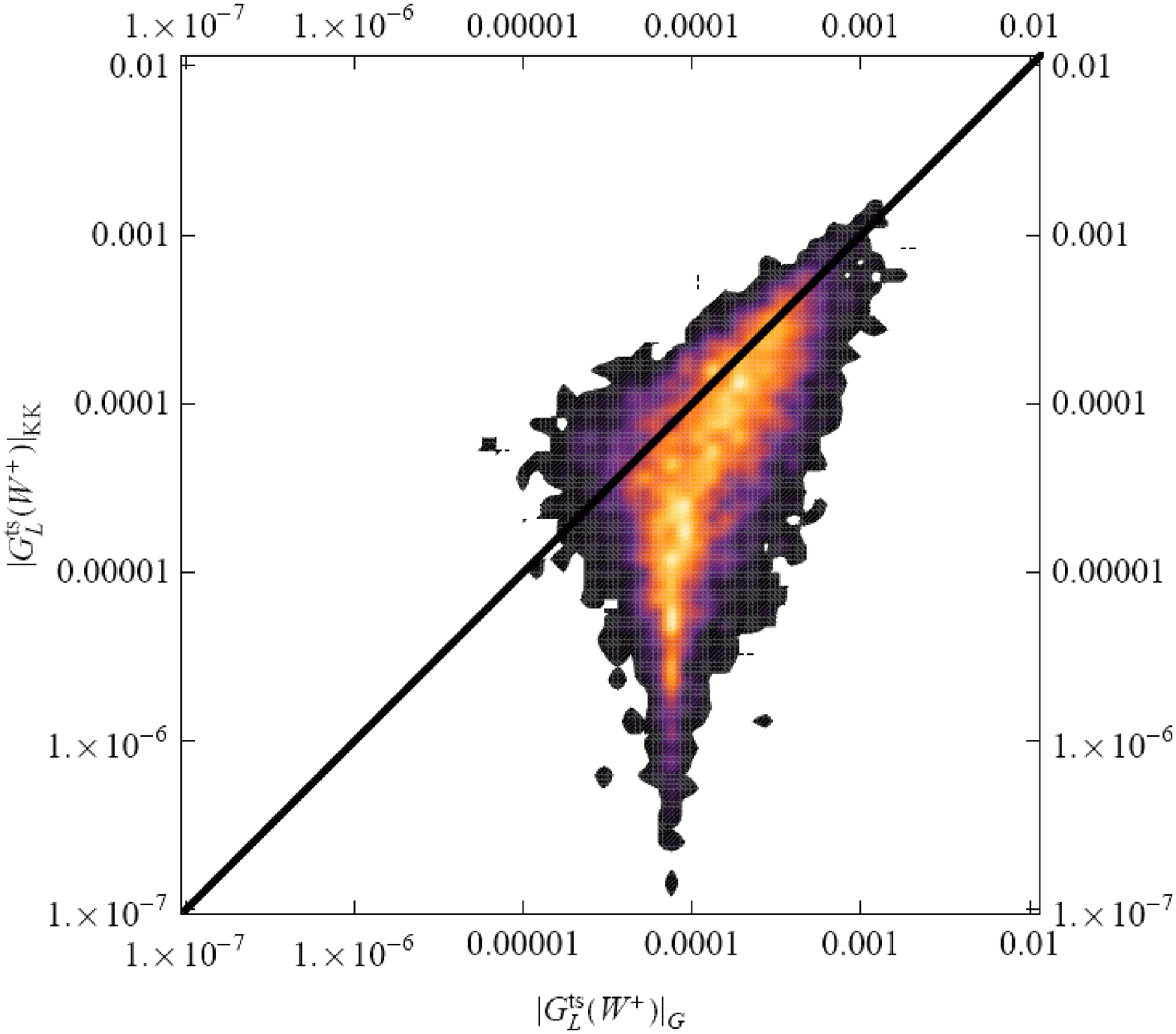}\hspace{.5cm}\includegraphics[width=.45\textwidth]{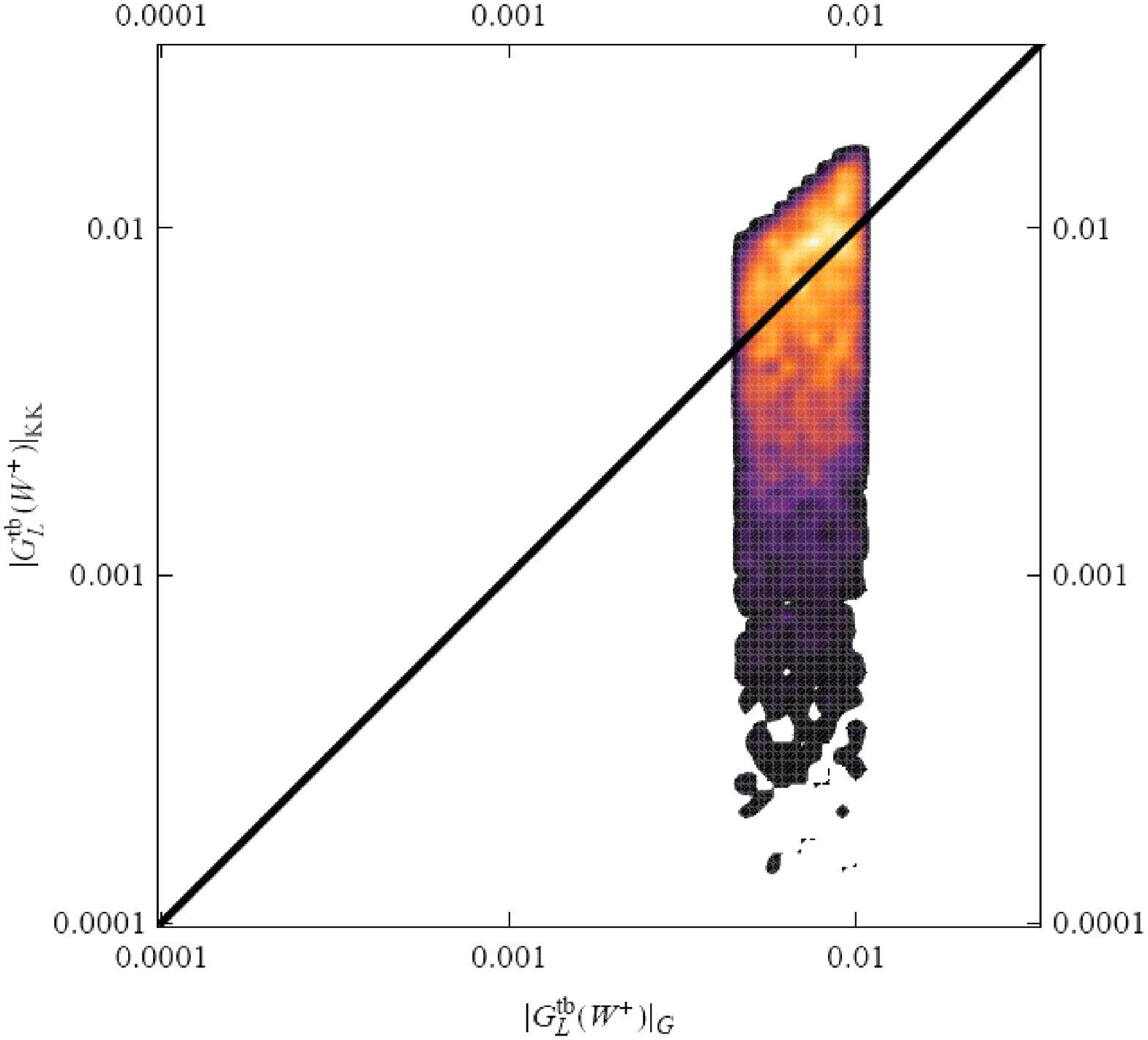}}
\caption{Comparison of contributions from KK fermion mixing and gauge boson mixing to the $W^+\bar t_L s_L$ coupling (left panel) and the $W^+\bar t_L b_L$ coupling (right panel).\label{fig:KK-contrib_W}}
\end{figure}

We next investigate the corrections to the elements of the CKM matrix due to the KK excitations of fermions and gauge bosons.
{From Sections~\ref{sec:nonuniversality} and \ref{sec:violationfermions} we expect that these corrections are of order $\mathcal{O}(v^2/M_\text{KK}^2)$.
In particular, if we compute the corrections due to KK fermions and KK gauge bosons separately and we define}
\begin{eqnarray}
\Delta V_\text{CKM}^\text{KK}&\equiv&\left|\frac{V_\text{CKM}^\text{KK}-V_\text{CKM}^0}{V_\text{CKM}^0}\right|\,,\\
\Delta V_\text{CKM}^\text{G}&\equiv&\left|\frac{V_\text{CKM}^\text{G}-V_\text{CKM}^0}{V_\text{CKM}^0}\right|\,,
\end{eqnarray}
{as} done in Section~\ref{sec:Size_of_Corrections_to_Z} for the $Z$ couplings we can also here predict the pattern of hierarchies in $\Delta V_\text{CKM}^\text{KK}$ and $\Delta V_\text{CKM}^\text{G}$. From (\ref{eq:G_L}) we can deduce that the hierarchy in $\Delta V_\text{CKM}^\text{G}$ is CKM-like, while the effective theory expressions imply that the pattern of $\Delta V_\text{CKM}^\text{KK}$ is the same as in the dyadic product $f_Q\circ f_Q$. A comparison of these patterns suggests that typically the largest impact of KK fermions relative to the effects of gauge boson mixing is expected in the $tb$ element, but also in the $ts$ and $cb$ elements. Both contributions to the $ts$ and $tb$ elements of the CKM matrix are compared to each other in Fig.~\ref{fig:KK-contrib_W}. Numerically we find
\be
\log_{10}{\Delta V_\text{CKM}^\text{KK}}\approx\begin{pmatrix}
-6.6_{-1.9}^{+1.9}&-4.9_{-1.6.0}^{+1.6}&-2.3_{-1.5}^{+1.0}\\-5.7_{-1.9}^{+1.6}&-4.9_{-1.5}^{+1.6}&-2.4_{-1.5}^{+0.7}\\-3.1_{-2.3}^{+1.2}&-3.0_{-2.0}^{+1.3}&-2.4_{-1.4}^{+0.6}\end{pmatrix}\,,
\ee
\be
\log_{10}{\Delta V_\text{CKM}^\text{G}}\approx\begin{pmatrix}
-2.8&-2.8_{-0.2}^{+0.0}&-2.1_{-1.0}^{+0.9}\\-2.8&-2.8_{-0.2}^{+0.0}&-2.2_{-0.9}^{+0.6}\\-2.6_{-1.0}^{+0.9}&-2.5_{-1.1}^{+0.9}&-2.2_{-0.1}^{+0.2}\end{pmatrix}\,,
\ee
where the quoted bounds (if given) enclose the values for $\Delta V_\text{CKM}^\text{KK}$, $\Delta V_\text{CKM}^\text{G}$ that are yielded by 99\% of all parameter points.
We observe that
\begin{itemize}
 \item as expected, the relative corrections are larger in the cases where the quarks of the third family are involved
 \item the $tb$ element of the CKM matrix is the only coupling in our analysis for which for a non-negligible portion of the parameter space the corrections coming from KK fermions can in principle be more important than the corrections from gauge boson mixing 
 \item for all other elements of the CKM matrix, even for $V_{ts}$ and $V_{cb}$, the corrections due to the KK gauge bosons are typically dominant.
\end{itemize}

\subsection{Coupling of $W^+$ to Right-Handed Quarks}
Following the line of argument in Section~\ref{sec:Size_of_Corrections_to_Z},
we find the counterpart of the CKM matrix describing the coupling of $W^+$ to
right-handed quarks to have a hierarchy among its elements that is very
different from that of the CKM matrix itself. Being characterised by the
pattern denoted by $f^u\circ f^d$ in Table~\ref{tab:hierarchies}, the elements
of $V_\text{R}$ increase mildly along its rows, e.g.~$V_\text{R}^{ud}<V_\text{R}^{us}<V_\text{R}^{ub}$ and more strongly along its columns, e.g. $V_\text{R}^{ud}<V_\text{R}^{cd}<V_\text{R}^{td}$. This implies that $V_\text{R}$ is neither approximately diagonal nor symmetric.

Numerically we find the entries of $V_\text{R}$ to have the typical values
\be
V_\text{R}\approx\begin{pmatrix}
1\cdot10^{-7}&1\cdot10^{-7}&3\cdot10^{-7}\\9\cdot10^{-6}&3\cdot10^{-5}&3\cdot10^{-5}\\8\cdot10^{-5}&2\cdot10^{-4}&9\cdot10^{-4}\end{pmatrix}\,
\ee
{for the scenario considered in \cite{Blanke:2008zb,Blanke:2008yr,Albrecht:2009xr}}

\newsection{Conclusions\label{sec:Conclusions}}
In the present paper we have derived general formulae for the corrections
to the SM gauge couplings to quarks resulting from the mixing of 
these quarks with the
heavy vector-like fermions in the process of EWSB. To this end we have integrated out the heavy fermions by means
of EOM. We have emphasised that in order to obtain correct 
results for the couplings in question
the kinetic terms of the SM fermions that are affected by this mixing
have to be brought into canonical form by a proper redefinition of these
fields.

We have applied our  formulae to the case of KK fermions 
in a RS
model with a custodial protection of flavour conserving $Zd_L^i\bar d_L^i$
and flavour violating $Zd_L^i\bar d_L^j$ couplings demonstrating explicitly
that this protection remains effective in the presence of the mixing with
KK fermions. In order to obtain this result
it was essential to bring the kinetic terms
of the SM fermions into the canonical form. Our numerical diagonalisation
of the fermion mass matrices gives another support to the correctness
of our formulae.

We have also pointed out that
in this
model the couplings of $Zu_R^i\bar u_R^i$ and $Zu_R^i\bar u_R^j$ are
also protected. {A list collecting the couplings that are protected in this particular RS model}
can be found in Section~\ref{sec:Custodial_protection_preliminaries}.

We have also shown explicitly that, in the limit discussed in section \ref{sec:Higgscouplings}, at $\mathcal{O}(v^2/M_\text{KK}^2)$ the
fermion--Higgs couplings of a Higgs placed on the IR brane is not
affected by the KK contributions up to strongly chirally suppressed
contributions discussed also in~\cite{Blanke:2008zb}.
 
An interesting implication of the mixing of the SM fermions with heavy 
vectorial fermions is the generation of the
 right-handed couplings of $W^\pm$. This mixing can be described by a
non-unitary $3\times 3$ matrix that describes the pattern of flavour
 violation in these new interactions of the standard $W^\pm$ boson.

 Our detailed numerical analysis results in the following findings:
\begin{itemize}
 \item The accuracy of the effective theory formulae that take into account mixing of SM quarks and KK fermions up to order $\mathcal O(v^2/M_\text{KK}^2)$ is excellent in the case of those $Z$ couplings that are not protected by the custodial symmetry and reasonably good in the case of $W^+$ couplings. The strong suppression of custodially suppressed couplings is qualitatively reproduced.
\item We find the impact of KK fermion mixing on custodially protected couplings to be smaller than the contribution from gauge boson mixing by at least five orders of magnitude. This explicitly shows that corrections from KK-fermion mixing do not spoil the custodial protection of the above couplings. 

 \item In the case of $Z$ and $W^\pm$ couplings that are not protected by the
 custodial symmetry we find the 
{KK fermion mixing effect} to be still smaller
 than the {gauge mixing effect} for a large majority of points in parameter space. This is in particular the case for those points in parameter space that produce large overall corrections to these couplings. The only exception to this rule is given by $(V_\text{CKM})_{tb}$ where corrections from KK fermion mixing and gauge boson mixing can have the same size.
 \item The violation of unitarity of the CKM matrix is smaller than 1\% for all unitarity relations except for the relations involving the third column and third row, respectively. Here unitarity can be violated by as much as 5\%. Also here the corrections from KK-fermion mixing typically have a smaller impact than the corrections stemming from gauge boson mixing, with the exception of the aforementioned unitarity relations involving the third column and third row where both corrections can be of the same size.
 \item Corrections to the CKM matrix elements themselves turn out to be as large as 2\% for $V_{tj}$, $j=d,s,b$, and $V_{ib}$, $i=u,c,t$, while corrections to all other elements are significantly smaller.

 \item The hierarchy in the 
 mixing matrix $V_{R}$ that is the analog of the CKM matrix for right-handed quarks is very much different from the $V_\text{CKM}$ one. In particular, we find that
 elements grow from left to right and top to bottom, that is $|V_{R}^{i+1,j}|>|V_{R}^{i,j}|$ and $|V_{R}^{i,j+1}|>|V_{R}^{i,j}|$, for $i,j=1,2,3$. Numerical values are found to range from $|V_{R}^{11}|\sim\mathcal O(10^{-7})$ to $|V_{R}^{33}|\sim\mathcal O(10^{-3})$ with intermediate values for the off-diagonal elements.

 \item In the limit discussed in section \ref{sec:Higgscouplings}, the off-diagonal non-derivative Higgs couplings {originating from $Y_1$ vanish} at order $\mathcal O(v^2/M_\text{KK}^2)$. {Terms involving derivatives, denoted by $Y_2$, however yield contributions to the off-diagonal Higgs couplings at this order.}

\end{itemize}

\subsubsection*{Acknowledgements}
\noindent We thank Monika Blanke, Csaba Csaki, Gino Isidori, Mikolaj Misiak and Andreas Weiler for useful discussions.
{Particular thanks go to Francisco del Aguila, Manuel Perez-Victoria and Jose Santiago for clarifying discussions related to their work in \cite{delAguila:2000kb,delAguila:2000aa,delAguila:2000rc}.}
AJB and BD thank the Cornell particle physics theory group for their warm hospitality at Cornell University during the final stages of this work.
This research was partially supported by {the Graduiertenkolleg GRK 1054, the Deutsche Forschungsgemeinschaft (DFG) under contract BU 706/2-1}, {the DFG Cluster of Excellence `Origin and Structure of the Universe' and by} the German Bundesministerium f{\"u}r Bildung und
Forschung under contract 05HT6WOA. {SG acknowledges support by the European Community's Marie Curie Research Training Network under contract MRTN-CT-2006-035505 [``HEPTOOLS''].}
\\

\begin{appendix}

\newsection{Couplings and Charge Factors}\label{Appendix}

In this Appendix we list all the couplings and the charge factors that we have 
used throughout this paper, and that can be easily computed using equations (\ref{eq:g}) and (\ref{eq:kappa}) together with Tables \ref{tab:fieldcontentSM} 
and \ref{tab:fieldcontent}. 

First, we give the charge factors in the couplings of SM down-quarks (both left 
and right-handed) to $Z$, $Z_X$ gauge bosons:

\begin{eqnarray}
g_{Z,L}^{4D}(d)=\frac{g^{4D}}{\cos\psi}\left[-\frac{1}{2}+\frac{1}{3}\sin^2\psi\right]\,,\label{eq:App1}\\
g_{Z,R}^{4D}(d)=\frac{g^{4D}}{\cos\psi}\left[\frac{1}{3}\sin^2\psi\right]\,,\label{eq:App2}\\
\kappa_1^{4D}(d)=\frac{g^{4D}}{\cos\phi}\left[-\frac{1}{2}-\frac{1}{6}\sin^2\phi
\right]\,,\\
\kappa_5^{4D}(d)=\frac{g^{4D}}{\cos\phi}\left[-1+\frac{1}{3}\sin^2\phi\right]\,.
\end{eqnarray}

Analogously, the charge factors in the couplings of SM up-quarks (both left and 
right-handed) to $Z$, $Z_X$ gauge bosons read:

\begin{eqnarray}
g_{Z,L}^{4D}(u)=\frac{g^{4D}}{\cos\psi}\left[\frac{1}{2}-\frac{2}{3}\sin^2\psi\right]\,,\\
g_{Z,R}^{4D}(u)=\frac{g^{4D}}{\cos\psi}\left[-\frac{2}{3}\sin^2\psi\right]\,,\\
\kappa_1^{4D}(u)=\frac{g^{4D}}{\cos\phi}\left[-\frac{1}{2}-\frac{1}{6}\sin^2\phi
\right]\,,\\
\kappa_3^{4D}(u)=\frac{g^{4D}}{\cos\phi}\left[-\frac{2}{3}\sin^2\phi\right]\,.
\end{eqnarray}

Finally, the charge factors in the couplings of the additional (vector-like) fermion fields 
($\chi^{u^i}$, $\chi^{d^i}$, $U^{\prime i}$, $U^{\prime\prime i}$ and 
$D^{\prime i}$) to $Z$, $Z_X$ gauge bosons are given by:

\bea
g_Z^{4D}\left(\chi^u\right)&=&\frac{g^{4D}}{\cos\psi}\left[\frac{1}{2}-\frac{5}{3}\sin^2\psi\right]\,,\\
 \kappa^{4D}\left(\chi^u\right)&=&\frac{g^{4D}}{\cos\phi}\left[\frac{1}{2}-\frac{7}{6
}\sin^2\phi\right]\,,\\
 g_{Z}^{4D}\left(\chi^d\right)&=&\frac{g^{4D}}{\cos\psi}\left[-\frac{1}{2}-\frac{2}{3
}\sin^2\psi\right]\\
\kappa^{4D}\left(\chi^d\right)&=&\frac{g^{4D}}{\cos\phi}\left[-\frac{1}{2}+\frac{5}{6
}\sin^2\phi\right]\,,\\
g_Z^{4D}\left(U^\prime\right)&=&g_Z(U^{\prime\prime})=\frac{g^{4D}}{\cos\psi}\left[-\frac{2}{3}\sin^2\psi\right]\,,\\
\kappa^{4D}\left(U^\prime\right)&=&\kappa\left(U^{\prime\prime}\right)=\frac{g^{4D}}{
\cos\phi}\left[-\frac{2}{3}\sin^2\phi\right]\,,\\
g_Z^{4D}\left(D^\prime\right)&=&\frac{g^{4D}}{\cos\psi}\left[-1+\frac{1}{3}\sin^2\psi
\right]\,,\label{eq:g_Z}\\
\kappa^{4D}\left(D^\prime\right)&=&\frac{g^{4D}}{\cos\psi}\left[\frac{4}{3}\sin^2\phi\right]\,.
\eea
\end{appendix}

\providecommand{\href}[2]{#2}\begingroup\raggedright\endgroup

\end{document}